\documentclass[letterpaper,twocolumn,amsmath,amssymb,showpacs,aps,prb,floatfix]{revtex4}
\usepackage[latin1]{inputenc}
\usepackage{color}
\usepackage{array}
\usepackage{float}
\usepackage{amsmath}
\usepackage{amssymb}
\usepackage{wasysym}
\usepackage{graphicx}
\usepackage{esint}
\usepackage{subscript}
\usepackage[unicode=true,pdfusetitle,
 bookmarks=true,bookmarksnumbered=false,bookmarksopen=false,
 breaklinks=false,pdfborder={0 0 1},backref=false,colorlinks=false]
 {hyperref}

\makeatletter

\pdfpageheight\paperheight
\pdfpagewidth\paperwidth

\providecommand{\tabularnewline}{\\}

\@ifundefined{textcolor}{}
{%
 \definecolor{BLACK}{gray}{0}
 \definecolor{WHITE}{gray}{1}
 \definecolor{RED}{rgb}{1,0,0}
 \definecolor{GREEN}{rgb}{0,1,0}
 \definecolor{BLUE}{rgb}{0,0,1}
 \definecolor{CYAN}{cmyk}{1,0,0,0}
 \definecolor{MAGENTA}{cmyk}{0,1,0,0}
 \definecolor{YELLOW}{cmyk}{0,0,1,0}
}

\renewcommand\Re{\operatorname{Re}}
\renewcommand\Im{\operatorname{Im}}

\makeatother

\begin{document}

\title{Scattering theory of spin--orbit active adatoms on graphene}

\author{Alexandre Pachoud,$^{1,2}$ Aires Ferreira$^{1,3}$, B. Özyilmaz$^{1,2,4}$
and A.~H. Castro Neto$^{1,2,5}$}

\affiliation{$^{1}$Graphene Research Centre and Department of Physics, National
University of Singapore, 2 Science Drive 3, Singapore 117546, Singapore}

\affiliation{$^{2}$NUS Graduate School for Integrative Sciences and Engineering
(NGS), National University of Singapore, Singapore 117456, Singapore}

\affiliation{$^{3}$Instituto de Física, Universidade Federal Fluminense, 24210-346
Niteroí, RJ, Brazil}

\affiliation{$^{4}$Nanocore, 4 Engineering Drive 3, National University of Singapore,
Singapore 117576}

\affiliation{$^{5}$Department of Physics, Boston University, 590 Commonwealth
Avenue, Boston, MA 02215, USA}
\begin{abstract}
The scattering of two-dimensional massless Dirac fermions from local
spin--orbit interactions with an origin in dilute concentrations of
physisorbed atomic species on graphene is theoretically investigated.
The hybridization between graphene and the adatoms' orbitals lifts
spin and valley degeneracies of the pristine host material, giving
rise to rich spin--orbit coupling mechanisms with features determined
by the exact adsorption position on the honeycomb lattice---bridge,
hollow, or top position---and the adatoms' outer-shell orbital type.
Effective graphene-only Hamiltonians are derived from symmetry considerations,
while a microscopic tight-binding approach connects effective low-energy
couplings and graphene--adatom hybridization parameters. Within the
$T$-matrix formalism, a theory for (spin-dependent) scattering events
involving graphene's charge carriers, and the spin--orbit active adatoms
is developed. Spin currents associated with intravalley and intervalley
scattering are found to tend to oppose each other. We establish that
under certain conditions, hollow-position adatoms give rise to the
spin Hall effect, through skew scattering, while top-position adatoms
induce transverse charge currents via trigonal potential scattering.
We also identify the critical Fermi energy range where the spin Hall
effect is dramatically enhanced, and the associated transverse spin
currents can be reversed.
\end{abstract}


\maketitle

\section{Introduction}

Graphene, being an atomically thin semi-metal, allows a fine control
over the charge carrier density via the electric-field effect.\cite{Novo04,AHCNReview}
Therefore, its basic characteristics---such as the density of states,\cite{Hobson53}
charge conductivity,\cite{Peres06,Ando06,Novo05} and electron-phonon
coupling\cite{AP10,Efetov10}---are gate-tunable, making this material
extremely versatile. The magnetic properties of graphene can be tailored
as well,\cite{Nair12} as placing the Fermi energy in the vicinity
of (away from) an adatom energy level switches on (off) magnetic moments.\cite{Uchoa08,Nair13} 

The properties of graphene can change even more drastically upon adsorption
of certain atomic species; highly conducting in its pristine form,
graphene can be transformed into an excellent insulator\cite{Nair10}
or a granular metal\cite{AP13} by chemisorption. Theoretical predictions
point out yet another possibility: the manipulation of spin--orbit
coupling (SOC) in chemically-modified graphene.\cite{AHCN-SOC09,weeks11,Gmitra_2013}
Such an appealing perspective has gained renewed interest recently
due to demonstrations of band-structure Rashba splitting in a graphene/Ni(111)
system with intercalated gold\cite{Marchenko12} and giant local SOC
enhancement in weakly hydrogenated graphene.\cite{Jaya} 

The control of SOC in gently modified graphene would open new directions
in carbon-based spintronics, complementing previous proposals requiring
magnetic fields\cite{Abanin} or magnetic ordering.\cite{Yazyev_Katsnelson}
In particular, \textcolor{black}{two-dimensional} (2D) Dirac fermions
in clean samples endowed with a SOC potential landscape could trigger
robust spin-polarized currents through the spin Hall effect (SHE)---a
phenomenon in which charge carriers with opposite spin states are
asymmetrically scattered under the action of a driving electric field.\cite{SHE}
\textcolor{black}{Given the importance of resonant scattering in 2D
carbon,}\cite{Scattering_Theory_SLG,Ferreira11}\textcolor{black}{{}
it is reasonable to expect that even dilute SOC-active ``hot spots''
would allow the observation of a sizable SHE. A recent study by some
of the authors predicted that nanometer-sized metal clusters provide
the required hot spots,\cite{SHE_G_14} capable of delivering giant
spin Hall coefficients reminiscent of those in pure metals.\cite{SHE_Exp_Metals}
Furthermore, the experimental data reported in Ref.~\onlinecite{Jaya}
strongly suggest that individual adatoms }\textcolor{black}{\emph{per
se}}\textcolor{black}{{} act as hot spots for enhanced spin-current
generation, although the underlying spin-transport mechanisms remain
unclear.\cite{Jaya,Gmitra_2013}}

In the present paper, the scattering of charge carriers from physisorbed
adatoms in monolayer graphene is investigated theoretically. The effective
SOC generated by generic adatoms is explicitly derived---within a
tight-binding and a continuum approach---and the associated single-impurity
problems are solved analytically using standard techniques. Regarding
the latter, our work generalizes the previous theory of resonant scattering
from short-range impurities\cite{Scattering_Theory_SLG,Ferreira11}
to situations in which the effective impurity potentials possess a
non-trivial spin and valley texture. Adatoms in their nonmagnetic
state are shown to mediate rich SOC scattering mechanisms---due to
an interplay between sublattice, spin, and valley degrees of freedom---that
cannot be observed in large clusters for which intervalley scattering
is inactive\textcolor{black}{.\cite{SHE_G_14}} It is already known
that $sp^{3}$ bonds formed by light adatoms induce large SOC due
to graphene lattice out-of-plane deformation\cite{AHCN-SOC09,Gmitra_2013}
with potential use in the SHE.\cite{Jaya} Here instead we investigate
the effects of heavy elements, which in a good approximation leave
the lattice flat. We show that such species can, in principle, induce
significant spin--orbit interactions if adsorbed in the hollow or
top position, but not in the bridge position. The actual scattering
mechanisms are seen to depend not only on the location of adatoms
on the lattice, but also on their valence orbital type, i.e., \emph{p},
\emph{d,} or \emph{f}. Strikingly, the competition between transverse
spin currents generated by intravalley and intervalley scattering
processes is shown to favor certain types of adatoms in the establishment
of SHE. Our calculations reveal that while hollow-position adatoms
can induce large pure transverse spin currents and SHE, top-position
adatoms produce transverse unpolarized currents via charge Hall effect
(CHE). 

The paper is organized as follows. Section~\ref{effective-adatom-Hamiltonian}
provides a derivation of effective low-energy SOC Hamiltonians induced
by generic atomic species physisorbed on graphene. This section generalizes
the particular case of \emph{p} outer-shell adatoms studied in Ref.~\onlinecite{weeks11}
to nonmagnetic adatoms with any type of valence orbital,\cite{Cohen_08,AdatomsGraphite,NonMag_Ni}
including the effects of intervalley scattering neglected in previous
studies. In addition to the case of hollow adatoms,\cite{weeks11}
we consider also the experimentally relevant cases of top and bridge
positions.\cite{Nair10,F_O} \textcolor{black}{The Hamiltonians derived
in this section describe dilute decorations, in which adsorbates located
at random positions act as scattering centers and the low-energy physics
is still dominated by graphene's $\pi$-electrons. The superlattice
dense limit, with adatoms occupying regularly spaced positions, endowing
the band structure with SOC, has been studied in Refs.~\onlinecite{weeks11,Gmitra_2013}.}

In Section~\ref{sec:Scattering-theory}, the scattering theory for
generic low-energy adatom models is established within a $T$-matrix
approach. In particular, explicit expressions for cross sections in
spin-conserving and spin-flip channels are given, and consequences
for the SHE and CHE are discussed. \textcolor{black}{The conditions
for the existence of low-energy regimes in which strong transverse
spin or charge currents arise from scattering} are discussed. Furthermore,
we show that, for certain adatoms in hollow and top positions, the
flow of transverse spin and charge currents can be reversed by tuning
the Fermi energy. 

Finally, the Appendix provides an alternative derivation of the effective
impurity Hamiltonians---derived in Section~\ref{effective-adatom-Hamiltonian}
using symmetry arguments---starting from microscopic tight-binding
Hamiltonians through Löwdin transformation. It also formulates conditions
for the appearance of large transverse spin and charge currents as
discussed in Section~\ref{sec:Scattering-theory} in terms of microscopic
hopping parameters.

\section{Effective adatom Hamiltonians\label{effective-adatom-Hamiltonian}}

In previous approaches in the continuum, adatoms have essentially
been modeled by a Dirac-peak potential $V\delta(\mathbf{r})$, in
order to estimate the charge dc-conductivity\cite{Peres06,Ferreira11}
of defective graphene, and to identify its dependence on the electronic
density. Although this approach proved successful experimentally,\cite{Manu11,Manchester10,Fuhrer09}
it does not capture subtle effects that we aim to investigate, in
particular the impact of adatoms on the charge carriers' spin and
valley degrees of freedom. In what follows, we establish a continuum
theory, within which an adatom situated at the origin adds a localized
effective-potential term $\mathcal{V}\delta(\mathbf{r})$ to pristine
graphene's Dirac Hamiltonian, and $\mathcal{V}$ is an $8\times8$
matrix which depends on the adatom's exact position in the lattice:
at the center of a honeycomb hexagon (hollow position), on top of
a carbon atom (top position), or in the middle of a carbon-carbon
bond (bridge position). For the continuum approach to incorporate
the most important symmetries associated with these particular adsorption
sites, we first derive very general graphene-only tight-binding Hamiltonians,
and eventually take the limit of vanishingly small lattice spacing.
The results derived here form the basis for the scattering theory
developed later in Sec.~\ref{sec:Scattering-theory}.

\subsection*{I.a Adatoms in hollow position}

\begin{figure}
\begin{centering}
\includegraphics[width=0.4\columnwidth]{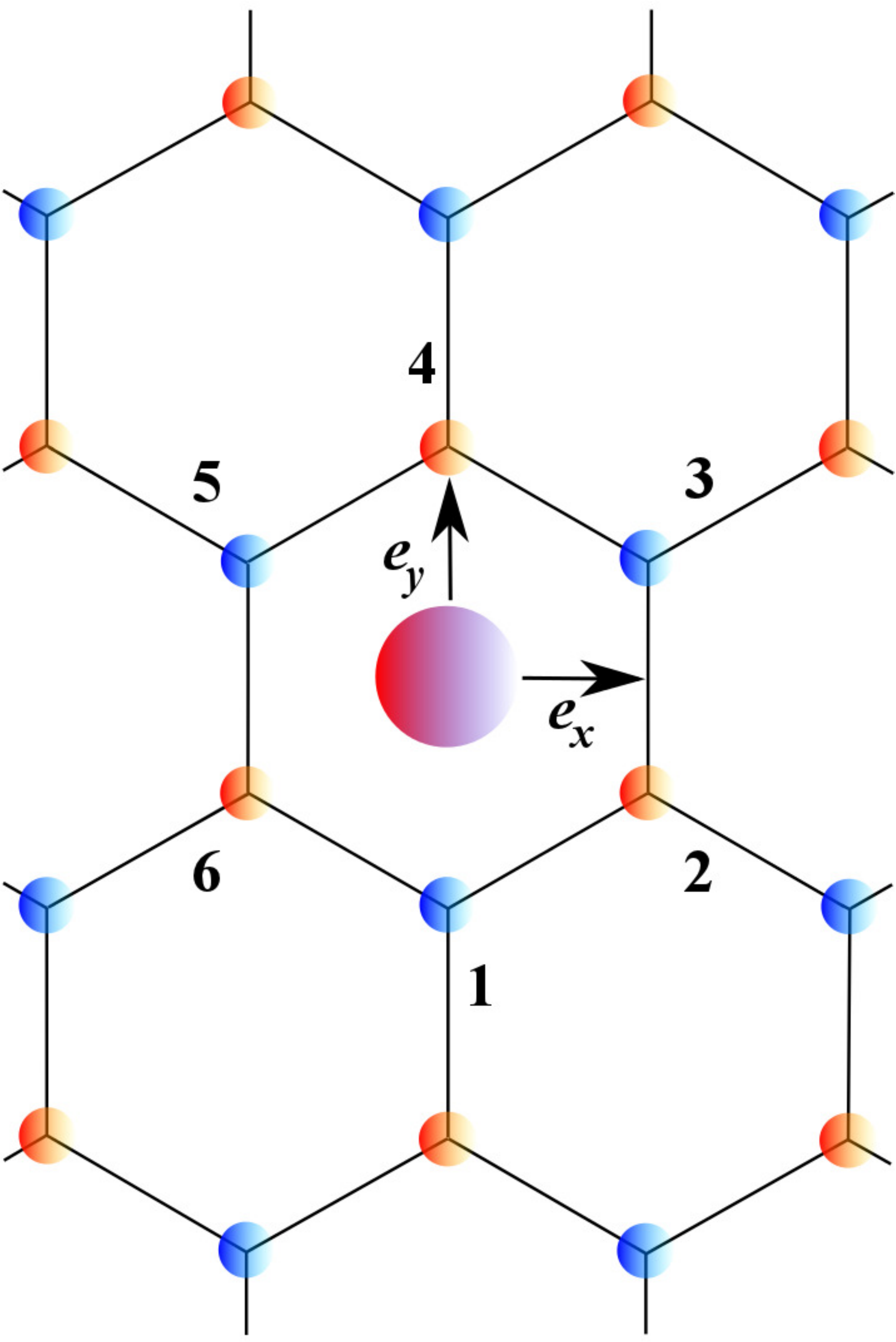}\,\includegraphics[width=0.6\columnwidth]{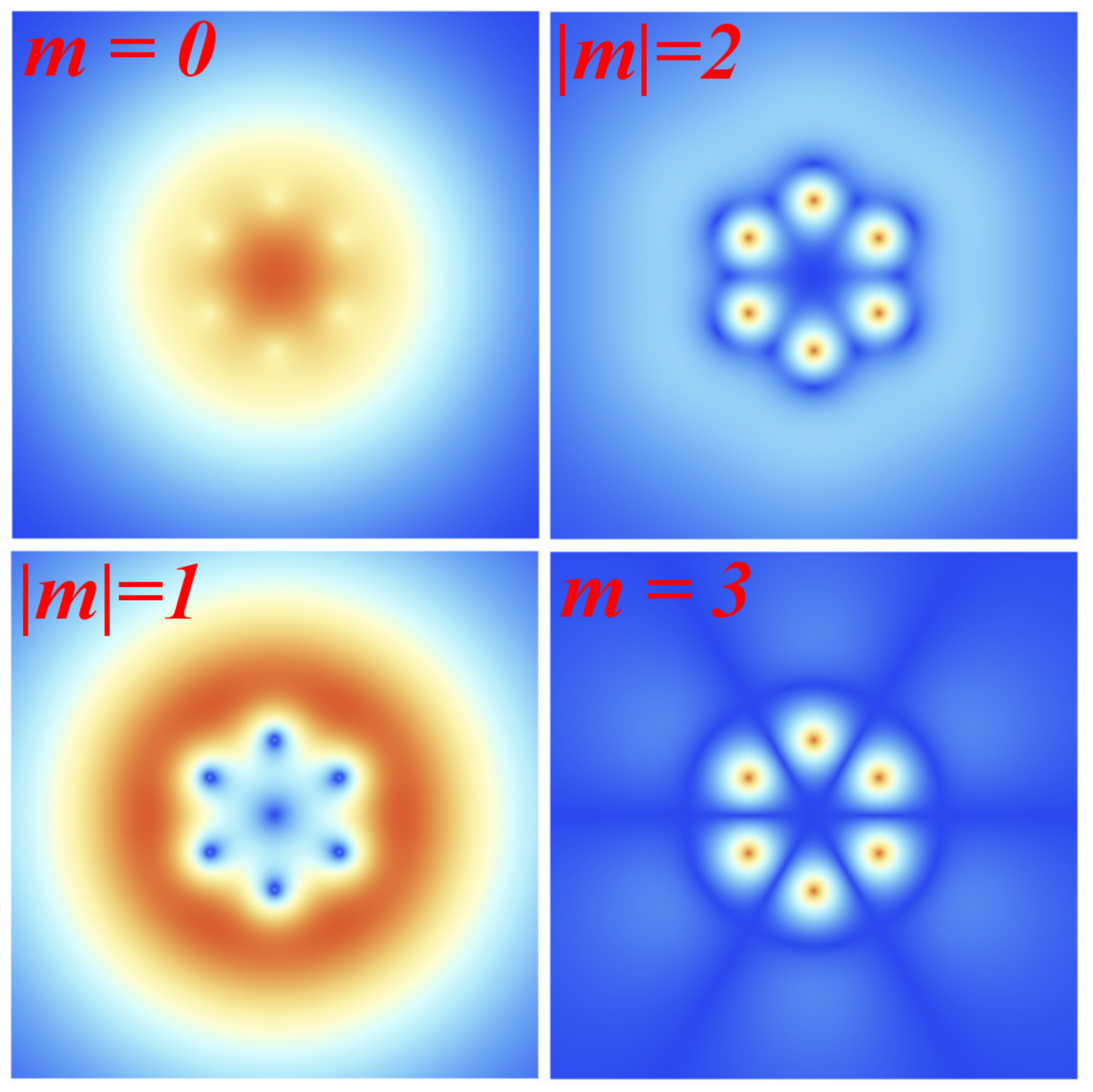}\linebreak{}

\par\end{centering}

\centering{}\caption{\label{fig:Hollow}Schematic picture of an adatom (pink sphere in
the left panel) in hollow position. $A$-sublattice ($B$-sublattice)
carbon atoms are represented as blue (red). Right panels show the
modulus of wave functions $\psi_{m}(x,y)$ created by operators $\Omega_{m}^{\dagger}$,
for $m=0,\pm1,\pm2,3$. Space coordinates $(x,y)$ have the adatom
as origin, and verify $(x,y)\in[-3.5,3.5]^{2}$ in units of $a\approx1.43\AA$.
The color scale is linear and represents $\left|\psi_{m}\right|$,
from dark blue (lowest values) to red (highest values). }
\end{figure}

We start by considering the case of a single adatom in the hollow
position. \textcolor{black}{As stated in the Introduction, our main
goal is to develop a general description of scattering from SOC-active
nonmagnetic atomic species weakly affecting carbon-carbon bonds.\cite{weeks11,Cohen_08}
Consequently, it suffices to describe the graphene--adatom system
with an effective $p_{z}$-orbital tight-binding model, supplemented
with a local interaction term describing the hybridization of graphene's
carbon atoms with the relevant adatom (outer-shell) orbital.}\textcolor{blue}{{}
}As we are primarily concerned with SOC, we decide to write our graphene-only
Hamiltonian in terms of creation and annihilation operators of states
with well-defined angular momentum, instead of the more conventional
creation and annihilation operators of carbon $p_{z}$-orbital states.
Since hopping integrals decrease exponentially with distance from
the adatom, the relevant states of definite angular momentum can be
written as a superposition of all $p_{z}$-orbital states located
at the six vertices of the hexagon occupied by the adatom, as depicted
in Fig.\,\ref{fig:Hollow}.

\begin{figure}
\begin{centering}
\includegraphics[width=0.7\columnwidth]{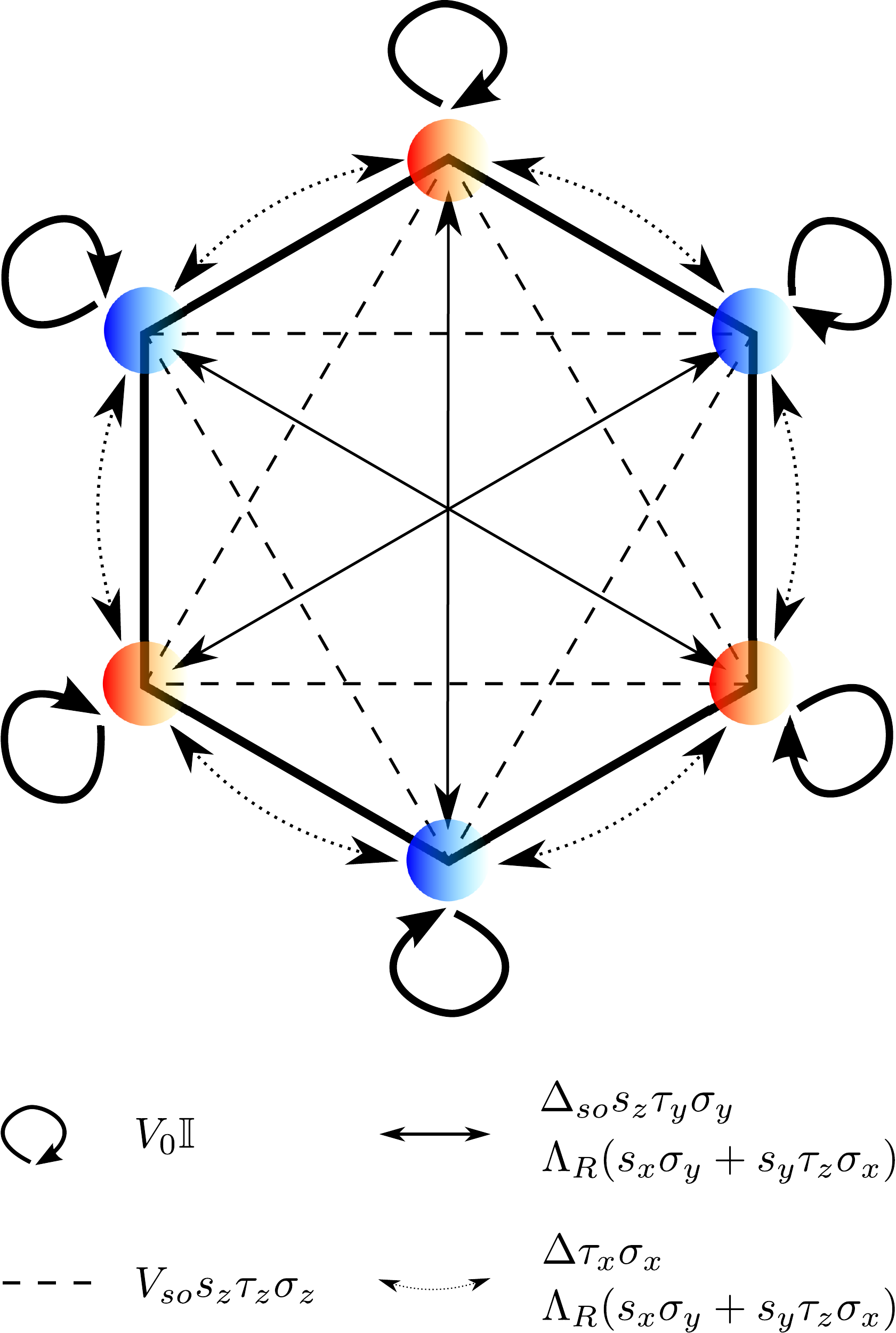}
\par\end{centering}

\caption{\label{fig:Hollow_Hopping}Interpretation of the effective Hamiltonian
$\mathcal{H}_{\textrm{hollow}}$ in terms of the hopping between graphene's
$p_{z}$ orbitals. Given the short-range nature of the adatom-induced
effective potential all atoms around the adatom mutually interact.
Intervalley terms arise from hopping between adjacent atoms. Although
they were neglected in previous works,\cite{weeks11,Gmitra_2013}
these terms play a central role in scattering from dilute random ensembles
of adatoms. }
\end{figure}

Using the numbering of carbon atoms shown in Fig.\,\ref{fig:Hollow},
and denoting the operator annihilating a $p_{z}$-orbital state of
atom $n$ by $c_{n}$ , we focus on operators of the form $C=\sum_{n=1}^{6}\lambda_{n}c_{n}$.
We also denote by $s_{l=x,y,z}$ the Pauli matrices acting on spin.
Requiring $C$ to have a given angular momentum, there exists an integer
$m$ such that $C$ transforms into $e^{-is_{z}\pi/6}e^{-im\pi/3}C$
under in-plane rotation by $\pi/3$ around the adatom. This condition
imposes that $\lambda_{2}=\omega^{m}\lambda_{1}$, ..., $\lambda_{6}=\omega^{5m}\lambda_{1}$,
where $\omega=e^{-i\pi/3}$, i.e., the \textit{\emph{only}} possible
operators annihilating a quasiparticle state on the hexagonal plaquette
hosting the adatom, and of well-defined angular momentum around this
adatom are, up to a scalar coefficient and unitary operator acting
on spin, $\Omega_{m}=\sum_{n=1}^{6}\omega^{m(n-1)}c_{n}$ for $m=0,\pm1,\pm2,3$.
By construction, operators $\Omega_{m}$, already encountered in Ref.~\onlinecite{weeks11},
carry angular momentum $m$, except $\Omega_{3}$ which carries angular
momentum 0. This can easily be seen by considering the time-reversed
operator $s_{y}\Omega_{3}$ which has the same angular momentum. Since
time-reversal transforms angular momentum $\mathbf{L}=\mathbf{r}\times\mathbf{p}$
into its opposite, it follows that $\Omega_{3}$ has angular momentum
zero. We also observe that the six operators $\Omega_{m}$ are linearly
independent, since $\left[\omega^{m(n-1)}\right]_{(m,n)}$ is a Vandermonde
matrix \cite{Vandermonde} and $\omega$ is a primitive sixth root
of unity. Therefore, the most general graphene-only single-electron
Hamiltonian term induced by a hollow-position adatom can be written
in terms of operators annihilating ``hexagonal'' $\Omega_{m}^{\dagger}|0\rangle$
states, that is

\begin{eqnarray}
H_{\textrm{hollow}} & = & \sum_{m=-2}^{3}\Omega_{m}^{\dagger}X_{m}\Omega_{m}+\sum_{m=-2}^{1}\Omega_{m}^{\dagger}M_{m}\Omega_{m+1}\nonumber \\
 & + & \sum_{m=0,\pm1}\Omega_{3}^{\dagger}T_{m}\Omega_{m}+\textrm{H.c.}\,,\label{eq:Hhollow}
\end{eqnarray}
where $X_{m}$, $M_{m},$ and $T_{m}$ are matrices acting on spin.
These matrices connect operators $\Omega_{i}$ that have angular momenta
differing by at most by the unity, by conservation of total angular
momentum $J=L+S$. Conservation of $J$ also implies that $X_{m}$
and $T_{0}$ are diagonal matrices, while $M_{m}$ and $T_{1}$ matrices
are proportional to the spin-raising operator $s_{+}=\frac{s_{x}+is_{y}}{2}$
and $T_{-1}$ is proportional to the spin-lowering operator $s_{-}=\frac{s_{x}-is_{y}}{2}$.
This means that under a rotation by $\pi/3$, $\Omega_{m}^{\dagger}X_{m}\Omega_{m}$
and $\Omega_{m}^{\dagger}M_{m}\Omega_{m+1}$ are invariant, while
$\Omega_{3}^{\dagger}T_{m}\Omega_{m}$ terms are odd. The invariance
of Eq.~(\ref{eq:Hhollow}) under rotations by $\pi/3$ thus requires
$T_{m=0,\pm1}=0$, i.e., zero-angular-momentum states $\Omega_{3}^{\dagger}|0\rangle$
do not couple to any other hexagonal $\Omega_{m}^{\dagger}|0\rangle$
state and can be ignored. Moreover, since we consider non-magnetic
and static impurities, $H_{\textrm{hollow}}$ is time-reversal invariant,
which further implies that $s_{y}X_{m}^{*}s_{y}=X_{-m}$ and $s_{y}M_{m}^{*}s_{y}=M_{-m-1}^{\dagger}$.
Finally, the invariance of $H_{\textrm{hollow}}$ under reflection
$x\mapsto-x$ requires $s_{x}X_{m}s_{x}=X_{-m}$ and $s_{x}M_{m}s_{x}=M_{-m-1}^{\dagger}$.
Altogether, these conditions lead to the simplified form

\begin{eqnarray}
H_{\textrm{hollow}} & = & \sum_{m=-2}^{2}\nu_{m}^{+}\Omega_{m}^{\dagger}\Omega_{m}+\sum_{m=\pm1,\pm2}\nu_{m}^{-}\Omega_{m}^{\dagger}s_{z}\Omega_{m}\nonumber \\
 & + & i\sum_{m=-2}^{1}\Lambda_{m}(\Omega_{m}^{\dagger}s_{+}\Omega_{m+1}-\Omega_{m+1}^{\dagger}s_{-}\Omega_{m})\,,\label{eq:Hhollow_simple}
\end{eqnarray}
where $\nu_{m}^{\pm},\Lambda_{m}\in\mathbb{R}$ satisfying $\nu_{m}^{\pm}=\pm\nu_{-m}^{\pm}$
and $\Lambda_{-m-1}=-\Lambda_{m}$. It is important to note that Eq.~(\ref{eq:Hhollow_simple})
is general at the single-electron level, provided that interactions
between the adatom and graphene's $p_{z}$ orbitals are negligible
outside the adatom's six nearest neighbors. The exact coupling mechanisms
between hexagonal states will only affect the value of constants $\nu_{m}^{\pm}$
and $\Lambda_{m}$, but not the overall form of $H_{\textrm{hollow}}$
given by Eq.~(\ref{eq:Hhollow_simple}). The relations between these
coupling constants and the actual microscopic parameters, such as
energy levels and SOCs of the adatom, are derived in the Appendix
in situations where a graphene electron or hole in state $\Omega_{m}^{\dagger}|0\rangle$
undergoes spin-dependent tunneling to an adatom orbital of same angular
momentum, potentially flips its spin by intra-atomic spin--orbit interaction,
and tunnels back to the graphene sheet in another $\Omega_{m'\neq m}^{\dagger}|0\rangle$
state. In such cases, $H_{\textrm{hollow}}$ describes the effect
of graphene--adatom hybridization on the low-energy massless Dirac
fermions.

We now derive an expression for $H_{\textrm{hollow}}$ in the continuum-limit,
where the carbon-carbon distance $a_{0}$ is seen as vanishingly small.
In this limit, pristine graphene's Hamiltonian acquires the familiar
form \cite{Semenoff} 
\begin{equation}
H_{0}=\int\frac{d^{2}\mathbf{r}}{A_{\hexagon}}\Psi^{\dagger}(\mathbf{r})\mathcal{H}_{0}(\mathbf{r})\Psi(\mathbf{r})\,,\label{eq:ContinuumH0}
\end{equation}
where $\mathcal{H}_{0}=v_{F}(\tau_{z}\sigma_{x}p_{x}+\sigma_{y}p_{y})$
and $A_{\hexagon}=3\sqrt{3}a_{0}^{2}/2$ is the unit cell area. Here,
$\tau_{l=x,y,z}$ and $\sigma_{l=x,y,z}$ are Pauli matrices acting,
respectively, on valley- and sublattice-space {[}for later use we
also define $\tau_{0},\sigma_{0},s_{0}\equiv\textrm{iden}(1,1)${]}.
In Eq.~\ref{eq:ContinuumH0},

\begin{eqnarray}
\Psi^{\dagger}(\mathbf{r})=(\Psi_{\uparrow KA}^{\dagger}(\mathbf{r}),\Psi_{\uparrow KB}^{\dagger}(\mathbf{r}),\Psi_{\uparrow K'A}^{\dagger}(\mathbf{r}),\Psi_{\uparrow K'B}^{\dagger}(\mathbf{r}),\,\,\nonumber \\
\Psi_{\downarrow KA}^{\dagger}(\mathbf{r}),\Psi_{\downarrow KB}^{\dagger}(\mathbf{r}),\Psi_{\downarrow K'A}^{\dagger}(\mathbf{r}),\Psi_{\downarrow K'B}^{\dagger}(\mathbf{r}))\,,\label{eq:spinor}
\end{eqnarray}
is an $8\times1$ creation operator whose components $\Psi_{s\tau\sigma}^{\dagger}(\mathbf{r})$
create a state at point $\mathbf{r}$ with spin $s=\,\uparrow,\downarrow\,\equiv1,-1$,
valley $\tau=K,K'\equiv1,-1$ and in sublattice $\sigma=A,B\equiv1,-1$.
To account for both $K$ and $K'$ valleys, we write spin-s components
of annihilation operators $c_{n}$ as superpositions of $\Psi_{sK\sigma_{n}}(\mathbf{r}_{n})$
and $\Psi_{sK'\sigma_{n}}(\mathbf{r}_{n})$, where $\mathbf{r}_{n}$
and $\sigma_{n}$ are the position vector and sublattice index corresponding
to site $n$, i.e.,
\begin{equation}
c_{n,s}=\sum_{\tau=\pm1}e^{i\tau\overrightarrow{\Gamma K}\cdot\mathbf{r}_{n}}\Psi_{s\tau\sigma_{n}}(\mathbf{r}_{n})\,.
\end{equation}
Here, $\Gamma$ denotes graphene's first Brillouin zone center, and
taking the $a_{0}\rightarrow0$ limit, the spin-$s$ component of
$\Omega_{m}$ becomes 
\begin{equation}
\Omega_{m,s}=\sum_{\tau}\left(\gamma_{m\tau}^{A}\Psi_{s\tau A}(\vec{0})+\gamma_{m\tau}^{B}\Psi_{s\tau B}(\vec{0})\right)\,,\label{eq:ContinuumOmega}
\end{equation}
with $\gamma_{m\tau}^{A}=1+2\cos[2\pi(m-\tau)/3]$ and $\gamma_{m\tau}^{B}=(-1)^{m}+2\cos[\pi(m-2\tau)/3]$.
Writing 
\begin{equation}
H_{\textrm{hollow}}=\int\frac{d^{2}\mathbf{r}}{A_{\hexagon}}\Psi^{\dagger}(\mathbf{r})\mathcal{H}_{\textrm{hollow}}(\mathbf{r})\Psi(\mathbf{r})\,,
\end{equation}
we obtain the following continuum-limit expression 
\begin{eqnarray}
\mathcal{H}_{\textrm{hollow}}(\mathbf{r}) & = & \left[V_{0}\mathbb{I}+\Delta\tau_{x}\sigma_{x}+V_{\textrm{so}}s_{z}\tau_{z}\sigma_{z}+\Delta_{\textrm{so}}s_{z}\tau_{y}\sigma_{y}\right.\nonumber \\
 &  & \left.+\Lambda_{R}\left(s_{x}\sigma_{y}+s_{y}\tau_{z}\sigma_{x}\right)\right]A_{\hexagon}\delta(\mathbf{r})\,,\label{eq:ContinuumHhollow}
\end{eqnarray}
where $\mathbb{I}$ is the 8$\times$8 identity matrix, $V_{0}=9(\nu_{1}^{+}+\nu_{2}^{+})$,
$\Delta=9(\nu_{2}^{+}-\nu_{1}^{+}),$ $V_{\textrm{so}}=9(\nu_{1}^{-}-\nu_{2}^{-})$,
$\Delta_{\textrm{so}}=9(\nu_{1}^{-}+\nu_{2}^{-})$ and $\Lambda_{R}=-9\Lambda_{1}$.
Hereafter, $A_{\hexagon}$ is set to unity, unless specified otherwise.
We remark that Eq.~(\ref{eq:ContinuumHhollow}) is only valid in
the vicinity of the Dirac point, as terms of order 1 or higher in
momentum $k$ have been neglected. It nonetheless gives insight regarding
possible SOC mechanisms induced by hybridization. Indeed, in addition
to expected on-site potential $V_{0}\mathbb{I}$\cite{Ferreira11}
and Kane-Mele\cite{KaneMele} intrinsic SOC terms $V_{\textrm{so}}s_{z}\tau_{z}\sigma_{z}$
discussed in Ref.~\onlinecite{weeks11}, $\mathcal{H}_{\textrm{hollow}}$
contains a spin-independent intervalley term $\Delta\tau_{x}\sigma_{x}$
and a term $\Delta_{\textrm{so}}s_{z}\tau_{y}\sigma_{y}$ which mixes
both spin and valley degrees of freedom. The presence of $\Delta\tau_{x}\sigma_{x}$
reflects the fact that atomically small impurities tend to act as
``white noise''\cite{Ando99} in momentum space and hence make intravalley
and intervalley scattering processes equiprobable. Similar to intrinsic
SOC term $V_{\textrm{so}}s_{z}\tau_{z}\sigma_{z}$, the term $\Delta_{\textrm{so}}s_{z}\tau_{y}\sigma_{y}$
 is even under $\mathcal{R}_{z}:z\mapsto-z$ reflection. However,
the former differs from the latter by its valley-connecting character,
itself a direct consequence of the short-range nature of adatoms.
Importantly, $\mathcal{H}_{\textrm{hollow}}$ also contains a term
$\mathcal{H}_{R}=\Lambda_{R}(s_{x}\sigma_{y}+s_{y}\tau_{z}\sigma_{x})\delta(\mathbf{r})$
originating from couplings between hexagonal states of total angular
momentum $J=\pm\frac{3}{2}$, namely $\Omega_{\pm1}^{\dagger}s_{\pm}\Omega_{\pm2}$
and $\Omega_{\pm2}^{\dagger}s_{\mp}\Omega_{\pm1}$ in Eq.~(\ref{eq:Hhollow_simple}).
Since \emph{p} orbitals accommodate two states of angular momentum
$\pm3/2$, \emph{p} outer-shell adatoms can in principle mediate spin--orbit
interactions between hexagonal states $\Omega_{1,\uparrow}^{\dagger}|0\rangle$
and $\Omega_{2,\downarrow}^{\dagger}|0\rangle$, thereby leading to
non-zero $\Lambda_{R}$. This is confirmed by calculations performed
with Löwdin's method, shown in the Appendix. However, $\Lambda_{\textrm{R}}$
should be significantly enhanced in situations where SOC is mediated
by \emph{d}- or \emph{f}-orbital adatoms, which host four states of
angular momentum $\pm3/2$.

The symmetries of $\mathcal{H}_{\textrm{R}}$ are interesting in their
own right. This term is odd under $\mathcal{R}_{z}$ but differs from
the familiar Bychkov-Rashba Hamiltonian $\mathcal{H}_{\textrm{BR}}=\Lambda_{\textrm{so}}(s_{x}\sigma_{y}-s_{y}\tau_{z}\sigma_{x})$
induced by out-of-plane electric fields,\cite{BychkovRashba} a possibility
already predicted in Ref.~\onlinecite{McCann12} for inversion symmetry-breaking
impurities. Similarly to the Bychkov-Rashba Hamiltonian, $\mathcal{H}_{R}$
is SO$(2)$-symmetric, as it should be for spin--orbit interactions
induced by hollow-position adatoms, which preserve graphene's $C_{6v}$
symmetry. However, we note that spinors $\psi(\mathbf{r})$ verifying
$(\mathcal{H}_{0}+\mathcal{H}_{\textrm{hollow}})\psi=E\psi$ transform
under rotation by $\phi$, denoted by $\mathcal{R_{\phi}}$, as $e^{+is_{z}\phi/2}e^{-i\tau_{z}\sigma_{z}\phi/2}\psi(\mathcal{R}_{-\phi}(\mathbf{r}))$.
As a result, $\tau\sigma-s$ is a conserved quantity, but \textit{not}
$\tau\sigma+s$. 

Finally, let us mention that the Hamiltonian $\mathcal{H}_{\textrm{hollow}}$
can easily be interpreted in terms of hopping between graphene's $p_{z}$
orbitals closest to the adatom, as illustrated by Fig.~\ref{fig:Hollow_Hopping}.
While scalar potential $V_{0}\mathbb{I}$ and intrinsic SOC term $V_{\textrm{so}}s_{z}\tau_{z}\sigma_{z}$
are associated with on-site energies and hopping between second-nearest
neighbors, intervalley terms $\Delta\tau_{x}\sigma_{x}$ and $\Delta_{\textrm{so}}s_{z}\tau_{y}\sigma_{y}$
correspond to first- and third-nearest neighbor hopping respectively.
In contrast with Bychkov-Rashba spin--orbit interaction $\mathcal{H}_{\textrm{BR}}$,
the $\Lambda_{R}(s_{x}\sigma_{y}+s_{y}\tau_{z}\sigma_{x})$ term is
associated with \emph{both }first- and third-nearest neighbor hopping.

\subsection*{I.b Adatoms in top position}

\begin{figure}[h]
\begin{centering}
\includegraphics[width=0.565\columnwidth]{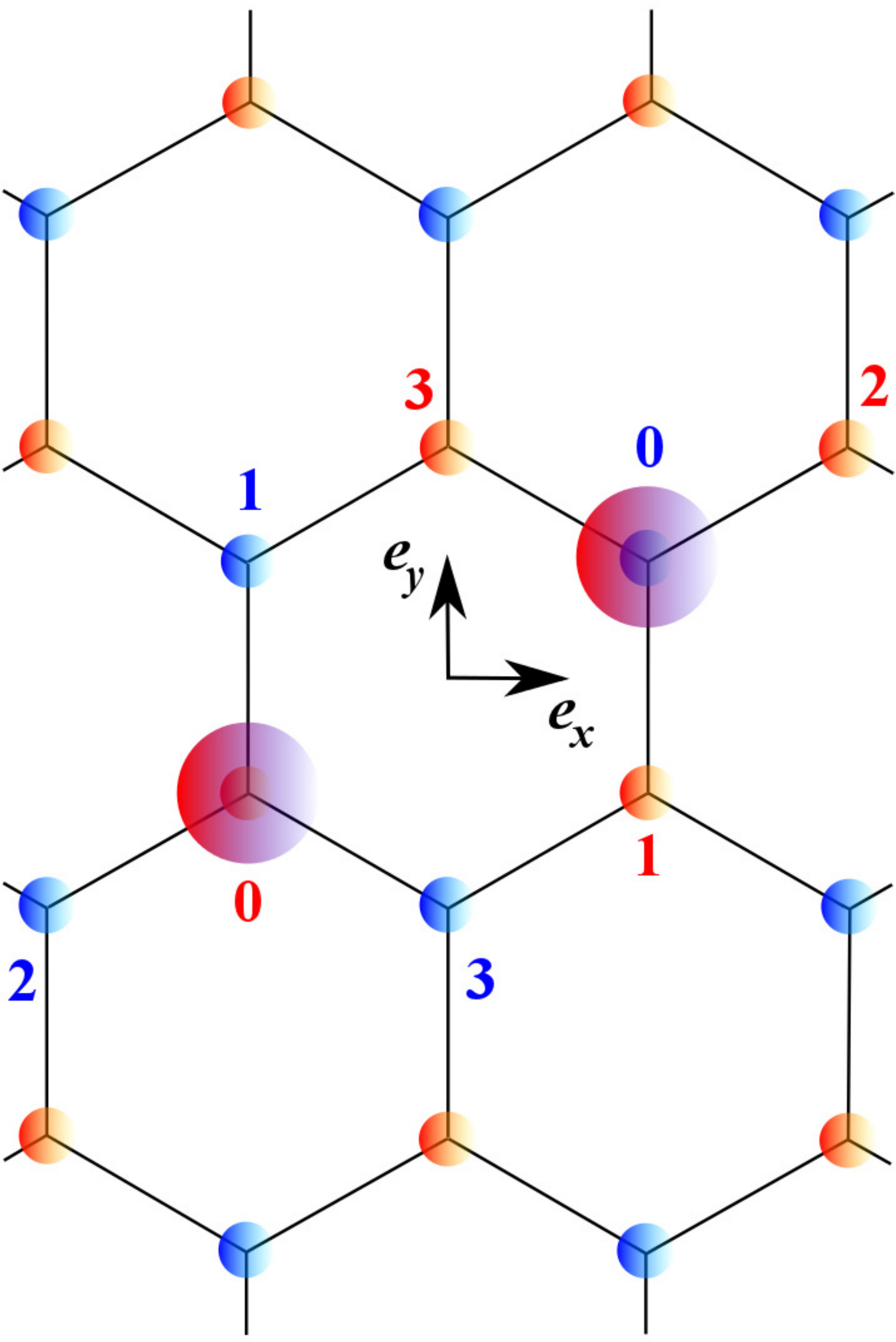}\,\includegraphics[width=0.435\columnwidth]{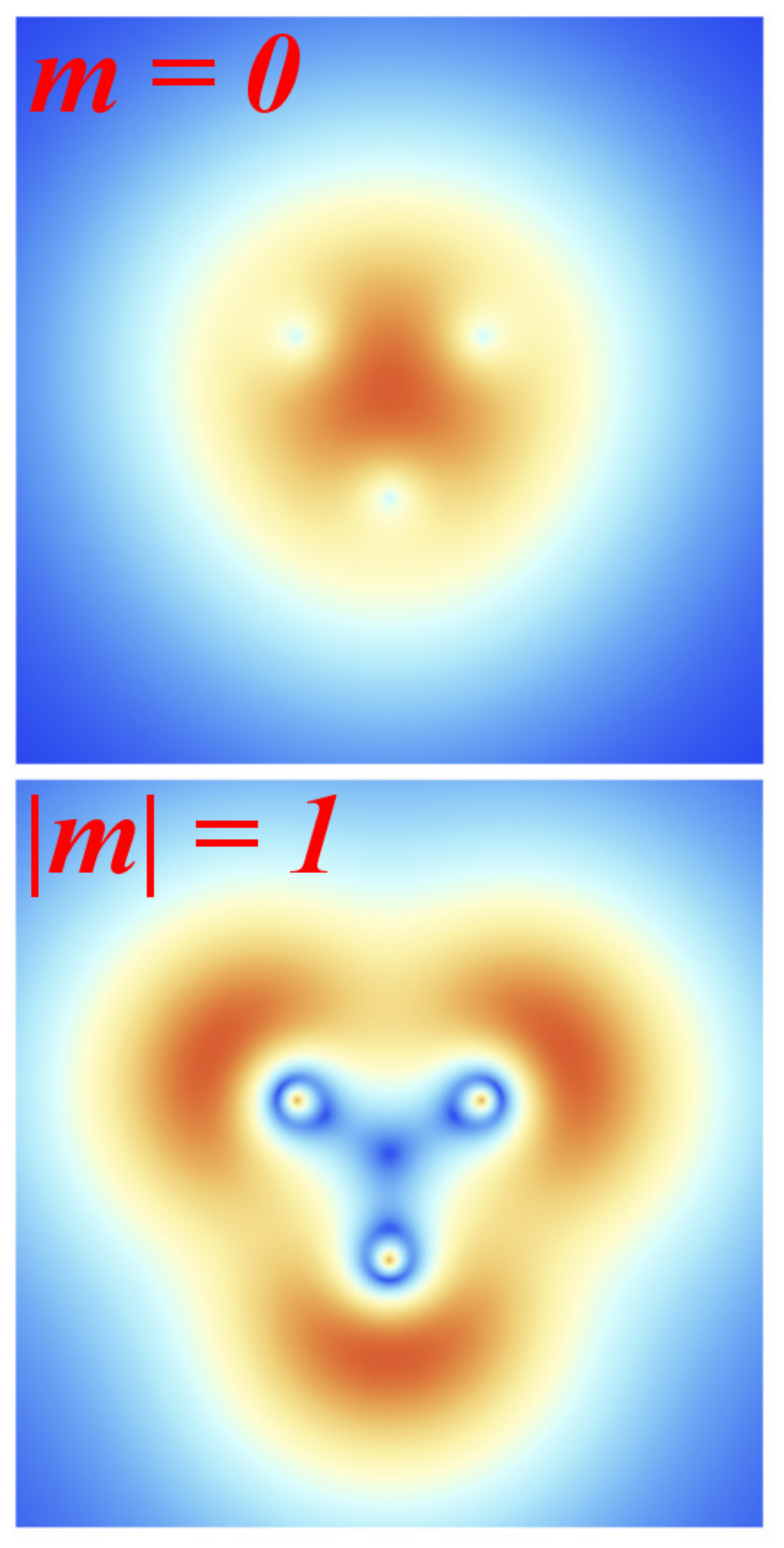}
\par\end{centering}

\caption{\label{fig:Top}Schematic picture of adatoms (pink spheres) in top
position, on an $A$-sublattice (blue) and B-sublattice (red) carbon
atom. Sites numbering used in main text is shown for both $A$ and
$B$ sublattice. Right panels show the modulus of wave functions $\phi_{m}(x,y)$
created around the $A$-sublattice adatom by operators $\Gamma_{m}^{\dagger}$,
for $m=0,\pm1$. Space coordinates $(x,y)$ have this adatom as origin,
and verify $(x,y)\in[-3.5,3.5]^{2}$ in units of $a\approx1.43\AA$.
The color scale is linear and reprsents $\left|\phi_{m}\right|$,
from dark blue (lowest values) to red (highest values).}
\end{figure}

\begin{figure}[h]
\begin{centering}
\includegraphics[width=0.85\columnwidth]{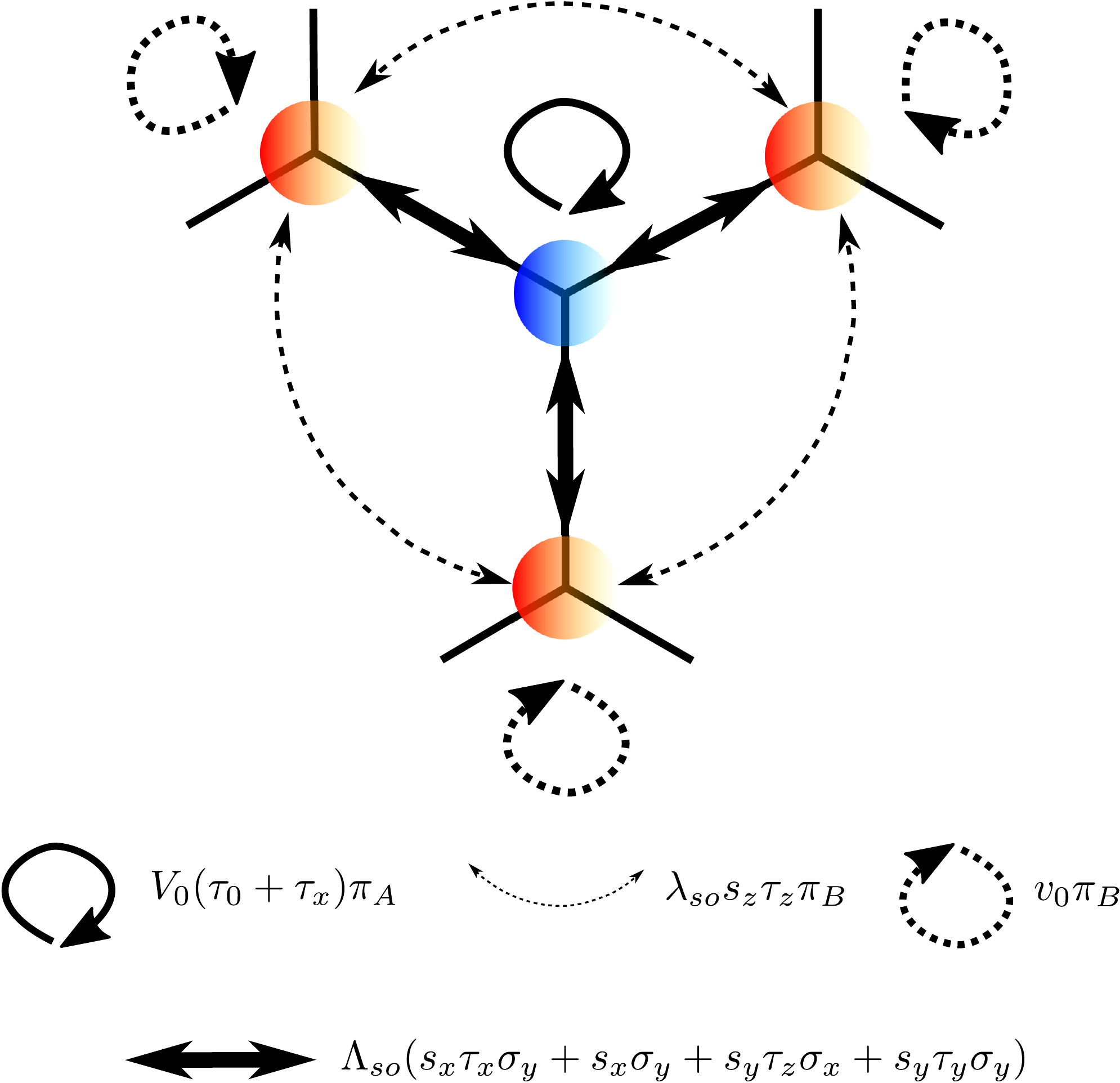}
\par\end{centering}

\caption{\label{fig:Top_Hopping}Interpretation of effective Hamiltonian $\mathcal{H}_{\textrm{top}}^{A}$
in terms of hopping between graphene's $p_{z}$ orbitals.}
\end{figure}
Another important class of adatoms are species that can be physisorbed
or chemisorbed in the top position, i.e., on top of a graphene carbon
atom belonging to the $A$- or $B$-sublattice, as depicted in Fig.~\ref{fig:Top}.
Such an adatom breaks graphene's $C_{6v}$ symmetry and hence induces
SOC mechanisms different from those introduced by adatoms in hollow
position; a top-position adatom has only one nearest neighbor, located
directly below and numbered as $0$ in Fig.~\ref{fig:Top} (as well
as three second nearest neighbors labeled 1,2 and 3 and situated at
a distance of $a_{0}$ away from 0). The electronic states with definite
angular momentum $m$ formed by linear combinations of $p_{z}$ orbitals
at sites 1,2 and 3 are ``triangular states'' annihilated by operators
$\Gamma_{m}=\sum_{n=1}^{3}\zeta^{m(n-1)}c_{n}$, where $m=0,\pm1$
and $\zeta=e^{-i2\pi/3}$. In particular, states with angular momentum
$\pm2$ are not supported. Since $\Gamma_{0}$, $\Gamma_{1}$ and
$\Gamma_{-1}$ are linearly independent, we can write the graphene-only
impurity Hamiltonian $H_{\textrm{top}}$, describing the action of
a top-position adatom, in terms of operators $c_{0}$ and $\Gamma_{0,\pm1}$
only, provided that interactions between the adatom and more distant
carbon atoms are negligible. The most general time-reversal invariant
$H_{\textrm{top}}$ conserving total angular momentum and preserving
$C_{3v}$-symmetry reads
\begin{eqnarray}
H_{\textrm{top}} & = & V_{0}c_{0}^{\dagger}c_{0}+V_{1}\Gamma_{0}^{\dagger}\Gamma_{0}+V_{2}\left(c_{0}^{\dagger}\Gamma_{0}+\Gamma_{0}^{\dagger}c_{0}\right)\nonumber \\
 & + & \Lambda_{+}\left(\Gamma_{1}^{\dagger}\Gamma_{1}+\Gamma_{-1}^{\dagger}\Gamma_{-1}\right)\nonumber \\
 & + & \Lambda_{-}\left(\Gamma_{1}^{\dagger}s_{z}\Gamma_{1}+\Gamma_{-1}^{\dagger}s_{z}\Gamma_{-1}\right)\nonumber \\
 & + & i\mu\left(c_{0}^{\dagger}s_{+}\Gamma_{1}+c_{0}^{\dagger}s_{-}\Gamma_{-1}-\textrm{H.c.}\right)\nonumber \\
 & + & i\tau\left(\Gamma_{0}^{\dagger}s_{+}\Gamma_{1}+\Gamma_{0}^{\dagger}s_{-}\Gamma_{-1}-\textrm{H.c.}\right)\,,\label{eq:Htop}
\end{eqnarray}
where $V_{0,1,2},\tau,\mu,\Lambda_{\pm}\in\mathbb{R}$. In the continuum-limit
$a_{0}\rightarrow0$, spin-\emph{s} components of $c_{0}$ and $\Gamma_{m}$
operators, are $c_{0,s}=\Psi_{sKA}(\vec{0})+\Psi_{sK'A}(\vec{0})$
and $\Gamma_{m,s}=3(1-\delta_{m,0})\Psi_{smB}(\vec{0})$ if the adatom
is on top of an $A$-sublattice carbon atom, and $c_{0,s}=\Psi_{sKB}(\vec{0})+\Psi_{sK'B}(\vec{0})$
and $\Gamma_{m,s}=3(1-\delta_{m,0})\Psi_{s,-m,A}(\vec{0})$, otherwise.
In the continuum, an adatom on top of an $A$,$B$-sublattice site
thus induces the following interaction
\begin{eqnarray}
\mathcal{H}_{\textrm{top}}^{A(B)}(\mathbf{r}) & = & \left[V_{0}\left(\tau_{0}+\tau_{x}\right)\pi_{A(B)}+v_{0}\pi_{B(A)}\right.\nonumber \\
 &  & \ \pm\lambda_{\textrm{so}}s_{z}\tau_{z}\pi_{B(A)}+\Lambda_{\textrm{so}}\left(s_{x}\tau_{x}\sigma_{y}\right.\notag\\
 &  & \left.\left.+s_{x}\sigma_{y}+s_{y}\tau_{z}\sigma_{x}\pm s_{y}\tau_{y}\sigma_{y}\right)\right]\delta(\mathbf{r})\,,\label{eq:ContinuumHtop}
\end{eqnarray}
where $\pi_{A(B)}=(\sigma_{0}\pm\sigma_{z})/2$ are projectors on
the $A$($B$)-sublattice, $v_{0}=9\Lambda_{+}$, $\lambda_{so}=9\Lambda_{-}$
and $\Lambda_{\textrm{so}}=\frac{9}{2}\mu$. The first term in Eq.~(\ref{eq:ContinuumHtop})
has already been derived for atomically sharp potentials.\cite{Ando99}
It induces intervalley scattering and is symmetric under $x\mapsto-x$
reflection $\mathcal{R}_{x}$, but breaks all rotational symmetries
in the continuum theory. The latter is borne out by the local density
of states maps in the vicinity of such impurities,\cite{Bena05} exhibiting
fringes perpendicular to $\overrightarrow{KK'}$ and hence to $\mathbf{e}_{x}$.
Invariance under $\mathcal{R}_{x}$ is manifest, as $\mathcal{U}_{x}H_{\textrm{top}}^{A(B)}\mathcal{U}_{x}=H_{\textrm{top}}^{A(B)}$,
where $\mathcal{U}_{x}=s_{x}\tau_{x}$ is the unitary representation
of $\mbox{\ensuremath{\mathcal{R}_{x}}}$ in the continuum theory
described by $\mathcal{H}=\mathcal{H}_{0}+\mathcal{H}_{\textrm{top}}^{A(B)}$.
Importantly, one also has $\mathcal{U}_{y}\mathcal{H}_{\textrm{top}}^{A}\mathcal{U}_{y}=\mathcal{H}_{\textrm{top}}^{B}$,
with $\mathcal{U}_{y}=s_{y}\sigma_{x}$, that is, $\mathcal{H}_{\textrm{top}}^{A}$
transforms into $\mathcal{H}_{\textrm{top}}^{B}$ under $\mathcal{R}_{y}:y\mapsto-y$,
faithfully reflecting the underlying lattice geometry. This means
that top-position adatoms induce different SOC terms, depending on
the host sublattice. Both close cousins of graphene's intrinsic SOC,
$\mathcal{R}_{z}$-even spin--orbit interaction mediated by top-position
adatoms on $A$- and $B$-sublattice are $\lambda_{\textrm{so}}s_{z}\tau_{z}\pi_{B}\delta(\mathbf{r})$
and $-\lambda_{\textrm{so}}s_{z}\tau_{z}\pi_{A}\delta(\mathbf{r})$,
respectively. The $\mathcal{R}_{z}$-odd component is more surprising.
In addition to the valley-preserving term $\propto(s_{x}\sigma_{y}+s_{y}\tau_{z}\sigma_{x})\delta(\mathbf{r})$
already encountered in Eq.~(\ref{eq:ContinuumHhollow}), a new valley-mixing
term $\Lambda_{\textrm{so}}(s_{x}\tau_{x}\sigma_{y}\pm s_{y}\tau_{y}\sigma_{y})\delta(\mathbf{r})$
emerges, $+$ for $\mathcal{H}_{\textrm{top}}^{A}$ and $-$ for $\mathcal{H}_{\textrm{top}}^{B}$.
Since in the continuum-limit $\Gamma_{0}=\mathcal{O}(a_{0}p)$, spin-flipping
processes coupling two triangular states are quenched, in contrast
with those coupling a triangular state $\Gamma_{\pm1}^{\dagger}|0\rangle$
with the central orbital $c_{0}^{\dagger}|0\rangle$, whose continuum-limit
is a superposition of $K$- and $K'$-valley states. This explains
why top-position adatoms give rise to $\mathcal{R}_{z}$-odd spin--orbit
interactions inducing both spin-flip and intervalley scattering. 

We finally note that, similarly to hollow-position adatoms, the continuum-limit
for top adatoms can be interpreted in terms of hopping between $p_{z}$
orbitals, as shown in Fig.~\ref{fig:Top_Hopping}. While spin-independent
terms $V_{0}(\tau_{0}+\tau_{x})\pi_{A(B)}$ and $v_{0}\pi_{B(A)}$
correspond to on-site energies on central site $0$ and neighboring
orbitals $i=1,2,3$, respectively, the $\mathcal{R}_{z}$-even SOC
term $\pm\lambda_{\textrm{so}}s_{z}\tau_{z}\pi_{B,A}$ is associated
with hopping between orbitals $i\ge1$. Finally, the $\mathcal{R}_{z}$-odd
term $\Lambda_{\textrm{so}}(s_{x}\tau_{x}\sigma_{y}+s_{x}\sigma_{y}+s_{y}\tau_{z}\sigma_{x}\pm s_{y}\tau_{y}\sigma_{y})$
arises from spin-dependent hopping between the central site $0$ and
its first nearest neighbors.

\subsection*{I.c. Adatoms in bridge position\label{sub:I.c.-Bridge} }

\begin{figure}[H]
\begin{centering}
\includegraphics[clip,width=0.55\columnwidth]{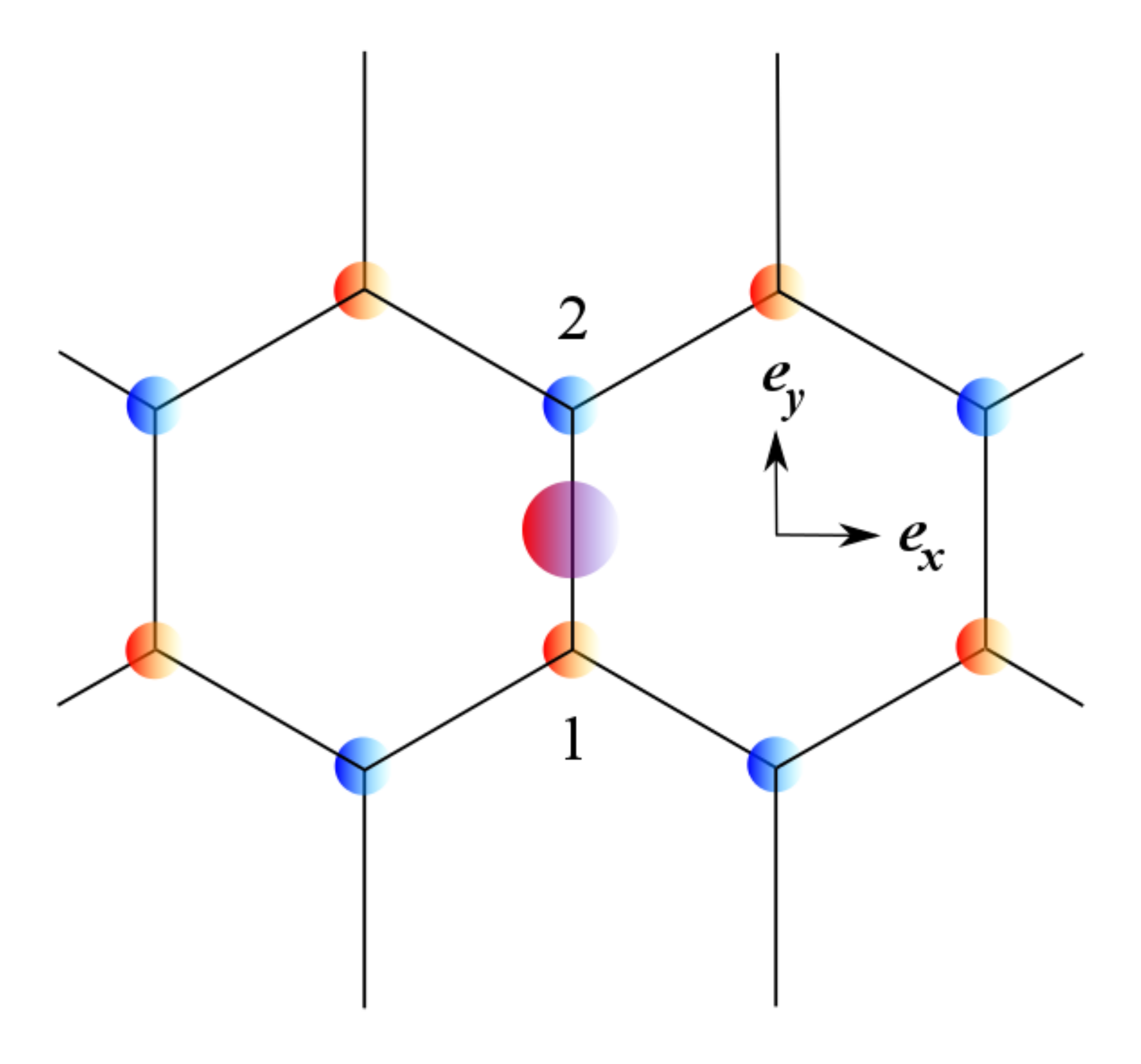}
\par\end{centering}

\caption{\label{fig:Bridge}Adatom (pink sphere) in the bridge position. $A$-
and $B$-sublattice carbon atoms are shown in blue and red, respectively.
Relevant atoms are numbered as in the main text.}
\end{figure}

We now consider the case of adatoms in the bridge position, depicted
in Fig.~\ref{fig:Bridge}. The only states of definite angular momentum
$m$ which can be formed with $p_{z}$ orbitals of atoms 1 and 2 are,
up to a scalar and a unitary matrix acting on spins: $(c_{1}^{\dagger}\pm c_{2}^{\dagger})|0\rangle$.
However, these states have angular momentum zero. Other possible linear
combinations of angular momentum $m$ including additional $p_{z}$
orbitals would also have $m=0$, because the only rotational symmetry
preserved by the bridge configuration is the rotation by $\pi$. As
a result, $\mathcal{R}_{z}$-odd SOC mechanisms induced by graphene--adatom
hybridization is forbidden in the absence of electric-field effects.
Furthermore, the impurity Hamiltonian $H_{\textrm{bridge}}$ induced
by any non-magnetic, static bridge-position adatom must respect hermiticity,
time-reversal symmetry, and $\mathcal{R}_{y}$. At the single electron
level, and limiting ourselves to orbitals 1 and 2, these conditions
constrain $H_{\textrm{bridge}}$ to read $H_{\textrm{bridge}}=V_{b}(c_{1}^{\dagger}c_{1}+c_{2}^{\dagger}c_{2})+\beta(c_{2}^{\dagger}c_{1}+c_{1}^{\dagger}c_{2})$,
where $V_{b},\beta\in\mathbb{R}$. Clearly, $H_{\textrm{bridge}}$
does \emph{not} have any SOC term, but for completeness, we derived
the continuum limit of $H_{\textrm{bridge}}$, using $c_{1,s}\rightarrow\Psi_{sKB}(\vec{0})+\Psi_{sK'B}(\vec{0})$
and $c_{2,s}\rightarrow\Psi_{sKA}(\vec{0})+\Psi_{sK'A}(\vec{0})$.
We obtain $\mathcal{H}_{\textrm{bridge}}=\left(\tau_{0}+\tau_{x}\right)\left(V_{b}+\beta\sigma_{x}\right)\delta(\mathbf{r})$.

\begin{widetext}

\begin{table}[H]
\begin{tabular}{|>{\centering}m{0.15\textwidth}|>{\centering}p{0.2\textwidth}|>{\centering}p{0.35\textwidth}|>{\centering}p{0.25\textwidth}|}
\hline 
~~~

Effective potential

~~~ & Hollow position & Top (A or B) position & Bridge position\tabularnewline
\hline 
\hline 
~~~

$\mathcal{V}_{\textrm{el}}$ & ~~~

$V_{0}\mathbb{I}+\Delta\tau_{x}\sigma_{x}$ & ~~~

$V_{0}(\tau_{0}+\tau_{x})\pi_{A(B)}+v_{0}\pi_{B(A)}$ & ~~~

$V_{b}(\tau_{0}+\tau_{x})+\beta(\tau_{0}+\tau_{x})\sigma_{x}$\tabularnewline
\hline 
~~~

$\mathcal{V}_{\textrm{even}}^{\textrm{so}}$ & ~~~

$V_{\textrm{so}}s_{z}\tau_{z}\sigma_{z}+\Delta_{\textrm{so}}s_{z}\tau_{y}\sigma_{y}$ & ~~~

$\pm\lambda_{\textrm{so}}s_{z}\tau_{z}\pi_{B(A)}$ & ~~~

$0$\tabularnewline
\hline 
~~~

$\mathcal{V}_{\textrm{odd}}^{\textrm{so}}$ & ~~~

$\Lambda_{R}(s_{x}\sigma_{y}+s_{y}\tau_{z}\sigma_{x})$ & ~~~

$\Lambda_{\textrm{so}}(s_{x}\tau_{x}\sigma_{y}+s_{x}\sigma_{y}+s_{y}\tau_{z}\sigma_{x}\pm s_{y}\tau_{y}\sigma_{y})$ & ~~~

$0$\tabularnewline
\hline 
\end{tabular}

\caption{\label{tab:EffectiveMass} Table comparing effective potentials induced
by spin-independent, $\mathcal{R}_{z}$-even and $\mathcal{R}_{z}$-odd
terms of impurity Hamiltonians originating from adatoms in hollow,
top (on $A$- or $B$-sublattice) and bridge positions. Results are
valid for adatoms with generic outer-shell orbital.}
\end{table}

\end{widetext}

In summary, adatoms in the hollow, top, and bridge positions give
rise to an interaction term $\mathcal{V}$ in the continuum-limit
Hamiltonian describing graphene quasi-particles,
\begin{equation}
\mathcal{H}(\mathbf{r})=v_{F}(\tau_{z}\sigma_{x}p_{x}+\sigma_{y}p_{y})+\mathcal{V}\delta(\mathbf{r})\,,
\end{equation}
where $\mathcal{V}=\mathcal{V}_{\textrm{el}}+\mathcal{V}_{\textrm{even}}^{\textrm{so}}+\mathcal{V}_{\textrm{odd}}^{\textrm{so}}$
are, in the close vicinity of the Dirac point, momentum-independent
$8\times8$ hermitian matrices. Here, $\mathcal{V}_{\textrm{el}}$
describes the spin-independent part (of pure electrostatic origin),
while $\mathcal{V}_{\textrm{even}}^{\textrm{so}}$ and $\mathcal{V}_{\textrm{odd}}^{\textrm{so}}$
correspond to $\mathcal{R}_{z}$-even and $\mathcal{R}_{z}$-odd SOC
contributions, respectively. Expressions for these matrices in the
previously discussed cases are compared in Table \ref{tab:EffectiveMass}.

\section{Scattering theory\label{sec:Scattering-theory}}

\subsection*{II.a. Scattering cross-section formalism}

In what follows, we consider a hollow- or top-position adatom on graphene,
centered at the origin and inducing an effective potential $\mathcal{V}\delta(\mathbf{r})$.
For concreteness, we take the spin-quantization axis along the $z$
axis (i.e., out-of-plane). The results derived below can be used to
obtain cross sections for arbitrary spin polarization following standard
recipes; see e.g., Ref.~\onlinecite{SHE_G_14}. The impurity induces
elastic scattering: incoming Dirac plane waves $\phi_{\mathbf{k}}^{s,\tau}(\mathbf{r})$
of spin $s$, valley $\tau$, momentum $\mathbf{k}$, and energy $E=s_{E}\hbar v_{F}k$,
where $s_{E}=\pm1=\textrm{sign}(E)$, scatter to outgoing waves $\phi_{\textrm{out}}(\mathbf{r})$,
giving rise to a total wave function $\Phi(\mathbf{r})=\phi_{\mathbf{k}}^{s,\tau}(\mathbf{r})+\phi_{\textrm{out}}(\mathbf{r})$
of energy $E$. The outgoing and incoming waves are related by the
the Lippmann-Schwinger equation\cite{LippmannSchwinger}
\begin{equation}
\phi_{\textrm{out}}(\mathbf{r})=G_{0}^{\pm}(\mathbf{r},E)T(E)\phi_{\mathbf{k}}^{s,\tau}(\vec{0})\,,
\end{equation}
where $G_{0}^{\pm}(\mathbf{r},E)=\langle\mathbf{r}|(E-\mathcal{H}_{0}+is_{E}0^{+})^{-1}|\vec{0}\rangle$
is the pristine graphene Green's function; here, $is_{E}0^{+}$ selects
the retarded or the advanced part, corresponding to outgoing waves
in electron and hole sectors, respectively. The matrix $T$ satisfies
the equation
\begin{equation}
T(E)=\mathcal{V}(\mathbb{I}-g(E)\mathcal{V})^{-1}\,.\label{eq:T}
\end{equation}
In the above, $g(E)$ is the retarded (advanced) Green's function
in the electron (hole) sector, evaluated at the origin, which is a
scalar given by
\begin{equation}
g(E)=\frac{E}{2\pi}\ln\frac{|E|}{E_{\textrm{c}}}-i\frac{E}{4}\,,\label{eq:g}
\end{equation}
where $E_{\textrm{c}}$ is graphene's half band-width and $\hbar v_{F}\equiv1$
has been set. For the sake of simplicity, we write the $T$~matrix
and $G_{0}^{\pm}(\mathbf{r},E)$ in the following basis of states
\begin{align}
\left\{ |\uparrow KA\rangle,|\uparrow KB\rangle,|\uparrow K'B\rangle,|\uparrow K'A\rangle,\right.\nonumber \\
\left.|\downarrow KA\rangle,|\downarrow KB\rangle,|\downarrow K'B\rangle,|\downarrow K'A\rangle\right\} ,\label{eq:MagicBasis}
\end{align}
where the Green's function takes the simple asymptotic form, as $r\rightarrow+\infty$,\cite{Ferreira11}
\begin{equation}
G_{0}^{\pm}(\mathbf{r},E)=-\sqrt{\frac{is_{E}k}{8\pi}}\frac{e^{ikr}}{\sqrt{r}}e^{i\tau_{z}\overrightarrow{\Gamma K}\cdot\mathbf{r}}(\mathbb{I}+s_{0}\tau_{z}\sigma_{\theta})\,,\label{eq:Green}
\end{equation}
with $\sigma_{\theta}=\cos\theta\sigma_{x}+\sin\theta\sigma_{y}$
and $\theta=\angle(\mathbf{e}_{x},\mathbf{r})$. In Eq.~(\ref{eq:Green}),
the diagonal matrix $e^{i\tau_{z}\overrightarrow{\Gamma K}\cdot\mathbf{r}}$
encodes the phase difference between waves at $K$ and $K'$ points.
In\emph{ }the basis defined by Eq.~(\ref{eq:MagicBasis}), we write
$T$ in block form:
\begin{equation}
T=\left(\begin{array}{cccc}
T_{\uparrow K,\uparrow K} & T_{\uparrow K',\uparrow K} & T_{\downarrow K,\uparrow K} & T_{\downarrow K',\uparrow K}\\
T_{\uparrow K,\uparrow K'} & T_{\uparrow K',\uparrow K'} & T_{\downarrow K,\uparrow K'} & T_{\downarrow K',\uparrow K'}\\
T_{\uparrow K,\downarrow K} & T_{\uparrow K',\downarrow K} & T_{\downarrow K,\downarrow K} & T_{\downarrow K',\downarrow K}\\
T_{\uparrow K,\downarrow K'} & T_{\uparrow K',\downarrow K'} & T_{\downarrow K,\downarrow K'} & T_{\downarrow K',\downarrow K'}
\end{array}\right)\,,
\end{equation}
and we denote, for valleys $\tau,\tau'$and spins $s,s'$, 
\begin{equation}
T_{s\tau,s'\tau'}=\left(\begin{array}{cc}
T_{s\tau,s'\tau'}^{11} & T_{s\tau,s'\tau'}^{12}\\
T_{s\tau,s'\tau'}^{21} & T_{s\tau,s'\tau'}^{22}
\end{array}\right)\,.
\end{equation}
The outgoing wave reads, away from the impurity,
\begin{equation}
\phi_{\textrm{out}}(\mathbf{r})=-\sqrt{\frac{is_{E}k}{8\pi}}\frac{e^{ikr}}{\sqrt{r}}e^{i\tau_{z}\overrightarrow{\Gamma K}\cdot\mathbf{r}}\sum_{s',\tau'}c_{s\tau,s'\tau'}(\theta)u_{k\mathbf{e}_{\mathbf{r}}}^{s',\tau'}(\theta)\,,\label{eq:Phiout}
\end{equation}
where $\mathbf{e}_{\mathbf{r}}=\mathbf{r}/r$, $u_{k\mathbf{e}_{\mathbf{r}}}^{s',\tau'}(\theta)=2^{-1/2}|s'\rangle\otimes|\tau'\rangle\otimes(1,s_{E}\tau'e^{i\theta})^{\textrm{t}}$
in\emph{ }basis (\ref{eq:MagicBasis}), and
\begin{eqnarray}
c_{s\tau,s'\tau'}(\theta) & = & T_{s\tau,s'\tau'}^{11}+\tau T_{s\tau,s'\tau'}^{12}\nonumber \\
 & + & \tau'e^{-i\theta}\left(T_{s\tau,s'\tau'}^{21}+\tau T_{s\tau,s'\tau'}^{22}\right)\,.\label{eq:c}
\end{eqnarray}
Accounting for both spin and valley degrees of freedom, the probability
density current associated with the outgoing wave reads
\begin{eqnarray}
\mathcal{J}_{r} & = & v_{F}\phi_{\textrm{out}}^{\dagger}s_{0}\tau_{z}\sigma_{\theta}\phi_{\textrm{out}}\nonumber \\
 & = & \frac{k}{8\pi r}v_{F}\sum_{s',\tau'}|c_{s\tau,s'\tau'}(\theta)|^{2}.\label{eq:probability_density_current}
\end{eqnarray}
The current associated with scattering of an incoming Dirac fermion
of spin $s$ and valley $\tau$ is thus the sum of currents $\mathcal{\mathcal{\boldsymbol{\mbox{\ensuremath{\mathcal{J}}}}}}_{s\tau,s'\tau'}=\frac{k}{8\pi r}v_{F}|c_{s\tau,s'\tau'}(\theta)|^{2}\mathbf{e}_{\mathbf{r}}$
arising from all possible $s\tau\rightarrow s'\tau'$ transitions,
and corresponding differential cross sections $\sigma_{s\tau,s'\tau'}(\theta)$
are 
\begin{equation}
\sigma_{s\tau,s'\tau'}(\theta)=\frac{k}{8\pi}\left|c_{s\tau,s'\tau'}(\theta)\right|^{2}\,.\label{eq:cross-section}
\end{equation}
An asymptotic formula for $\sigma_{s\tau,s'\tau'}(\theta)$ can easily
be derived from Eq.~(\ref{eq:c}): 
\begin{equation}
\sigma_{s\tau,s'\tau'}(\theta)=\frac{k}{8\pi}\left[\mathcal{C}_{s\tau,s'\tau'}^{2}+\mathcal{M}_{s\tau,s'\tau'}\cos(\theta+\varphi_{s\tau,s'\tau'})\right]\,,\label{eq:General_cross-section}
\end{equation}
where 
\begin{eqnarray}
\mathcal{C}_{s\tau,s'\tau'}^{2} & = & |T_{s\tau,s'\tau'}^{11}+\tau T_{s\tau,s'\tau'}^{12}|^{2}\nonumber \\
 & + & |T_{s\tau,s'\tau'}^{21}+\tau T_{s\tau,s'\tau'}^{22}|^{2}\,,\label{eq:Csquare}
\end{eqnarray}
\begin{eqnarray}
\mathcal{M}_{s\tau,s'\tau'} & = & 2\tau'|T_{s\tau,s'\tau'}^{11}+\tau T_{s\tau,s'\tau'}^{12}|\nonumber \\
 & \times & |T_{s\tau,s'\tau'}^{21}+\tau T_{s\tau,s'\tau'}^{22}|\,,\label{eq:M}
\end{eqnarray}
and
\begin{eqnarray}
\varphi_{s\tau,s'\tau'} & = & \arg(T_{s\tau,s'\tau'}^{11}+\tau T_{s\tau,s'\tau'}^{12})\nonumber \\
 & - & \arg(T_{s\tau,s'\tau'}^{21}+\tau T_{s\tau,s'\tau'}^{22})\,.\label{eq:varphi}
\end{eqnarray}
It is important to note that $\sigma_{s\tau,s'\tau'}(\theta)$ generally
has a phase $\varphi_{s\tau,s'\tau'}$ {[}see Eq.~(\ref{eq:General_cross-section}){]},
which can give rise to skew-scattering and thus SHE, provided $\varphi_{s\tau,s'\tau'}\neq0$
and $\mathcal{M}_{s\tau,s'\tau'}\neq0$. Establishing conditions under
which skew scattering is significant is the object of subsequent paragraphs.
This study can be conveniently carried out by comparing the integrated
skew cross-sections 
\begin{equation}
\Sigma_{s\tau,s'\tau'}^{\perp}=\int_{0}^{2\pi}d\theta\sin\theta\sigma_{s\tau,s'\tau'}(\theta),
\end{equation}
which measure the skewness of $s\tau\rightarrow s'\tau'$ scattering
mechanisms, to the integrated transport cross-sections
\begin{equation}
\Sigma_{s\tau,s'\tau'}^{\parallel}=\int_{0}^{2\pi}d\theta(1-\cos\theta)\sigma_{s\tau,s'\tau'}(\theta)\,.
\end{equation}
From a semi-classical view point, these integrated cross-sections
relate to the microscopic currents according to
\begin{align}
J_{s\tau,s'\tau'}^{\perp} & =s_{E}\int_{0}^{2\pi}\mathcal{\mathcal{\boldsymbol{\mbox{\ensuremath{\mathcal{J}}}}}}_{s\tau,s'\tau'}\cdot\mathbf{e}_{y}rd\theta\nonumber \\
 & =s_{E}v_{F}\Sigma_{s\tau,s'\tau'}^{\perp}\,,\label{eq:skew_micro_current}
\end{align}
and 
\begin{align}
J_{s\tau,s'\tau'}^{\parallel} & =s_{E}\int_{0}^{2\pi}\mathcal{\mathcal{\boldsymbol{\mbox{\ensuremath{\mathcal{J}}}}}}_{s\tau,s'\tau'}\cdot\left(\mathbf{e}_{\mathbf{r}}-\mathbf{e}_{x}\right)rd\theta\nonumber \\
 & =s_{E}v_{F}\Sigma_{s\tau,s'\tau'}^{\parallel}\,,\label{eq:long_micro_current}
\end{align}
associated with $s\tau\rightarrow s'\tau'$ processes. The knowledge
of \emph{microscopic} currents (\ref{eq:skew_micro_current})--(\ref{eq:long_micro_current}),
describing charge scattering with \emph{a} \emph{single} \emph{impurity},
gives direct access to the steady-state charge and spin \emph{macroscopic}
currents for a random ensemble of adatoms via the standard Boltzmann
transport formalism.\cite{SHE_G_14,SinovaBTE09}

\subsection*{II.b. Scattering with hollow-position adatoms}

We now focus on scattering mechanisms induced by an adatom in the
hollow position. Making use of Table \ref{tab:EffectiveMass}, the
calculated $T$~matrix in the basis given by Eq.~(\ref{eq:MagicBasis})
reads as
\begin{equation}
T_{\textrm{hollow}}=\left(\begin{array}{cc}
T_{\textrm{hollow}}^{\uparrow\uparrow} & T_{\textrm{hollow}}^{\downarrow\uparrow}\\
T_{\textrm{hollow}}^{\uparrow\downarrow} & T_{\textrm{hollow}}^{\downarrow\downarrow}
\end{array}\right)\,,
\end{equation}
with $4\times4$ blocks,
\begin{equation}
T_{\textrm{hollow}}^{ss}=\left(\begin{array}{cccc}
\alpha_{s} & 0 & \gamma_{s} & 0\\
0 & \beta_{s} & 0 & \delta_{s}\\
\gamma_{s} & 0 & \alpha_{s} & 0\\
0 & \delta_{s} & 0 & \beta_{s}
\end{array}\right)\,,\label{eq:Thollow1}
\end{equation}
where $s=\uparrow,\downarrow$ and 
\begin{equation}
T_{\textrm{hollow}}^{\uparrow\downarrow}=-[T_{\textrm{hollow}}^{\downarrow\uparrow}]^{\textrm{t}}=\left(\begin{array}{cccc}
0 & 0 & 0 & 0\\
\tau_{\textrm{f}} & 0 & 0 & 0\\
0 & 0 & 0 & 0\\
0 & 0 & -\tau_{\textrm{f}} & 0
\end{array}\right)\,.\label{eq:Thollow2}
\end{equation}
Matrix elements appearing in Eqs.~(\ref{eq:Thollow1})-(\ref{eq:Thollow2})
verify 
\begin{eqnarray}
\tau_{\textrm{f}} & = & \frac{2i\Lambda_{R}}{d}\,,\label{eq:Spin-flip_elements}
\end{eqnarray}
\begin{eqnarray}
\alpha_{\uparrow(\downarrow)} & =\beta_{\downarrow(\uparrow)}= & \frac{\chi_{\pm}-p_{\pm}g}{d_{\pm}}\,,\label{eq:puremom_1}
\end{eqnarray}
\begin{eqnarray}
\gamma_{\uparrow(\downarrow)} & =\delta_{\downarrow(\uparrow)}= & \frac{\Delta\mp\Delta_{\textrm{so}}}{d_{\pm}}\,,\label{eq:intervalley_elements-}
\end{eqnarray}
where we have set $\chi_{\varsigma}=V_{0}+\varsigma V_{\textrm{so}}$
and $p_{\varsigma}=(V_{0}+\varsigma V_{\textrm{so}})^{2}-(\Delta-\varsigma\Delta_{\textrm{so}})^{2}-2(1+\varsigma)\Lambda_{R}^{2}$,
with $\varsigma=\pm1$. We also defined
\begin{equation}
d_{\varsigma}=1-2g\chi_{\varsigma}+p_{\varsigma}g^{2}\,.\label{eq:d_pm}
\end{equation}
Strikingly, the $T$~matrix elements for intervalley scattering events
involving spin-flip are null. However, intravalley spin-flips, spin-preserving
intervalley scattering and pure momentum scattering events are allowed,
and we shall describe them in more detail in what follows.

Using Eqs.~(\ref{eq:c}) and (\ref{eq:cross-section}), we found
that differential cross-sections $\sigma_{\uparrow K,\downarrow K}(\theta)$,
$\sigma_{\uparrow K',\downarrow K'}(\theta)$, $\sigma_{\downarrow K,\uparrow K}(\theta)$,
and $\sigma_{\downarrow K',\uparrow K'}(\theta)$ are equal and isotropic,
leading to
\begin{equation}
\Sigma_{s\tau,-s\tau}^{\parallel}=\frac{k|\tau_{\textrm{f}}|^{2}}{4},
\end{equation}
and null skew cross-sections
\begin{equation}
\Sigma_{s\tau,-s\tau}^{\perp}=0\,,\label{eq:skew_cross_sec_spinflip}
\end{equation}
that is, spin-flip does not give rise to transverse spin currents.
We remark that a similar result has been recently derived for the
particular case of SO$(2)$-symmetric intravalley potentials,\cite{SHE_G_14}
which, in the present context, corresponds to setting $\Delta,\Delta_{\textrm{so}}=0$
in Eq.~(\ref{eq:ContinuumHhollow}). 

We now move gears to elastic (spin-preserving) scattering. In particular,
intervalley scattering cross-sections are characterized, for $\tau\neq\tau'$,
by
\begin{equation}
\mathcal{M}_{s\tau,s\tau'}=\tau'\mathcal{M}_{\textrm{inter}}\,,\label{eq:Intervalley_Hollow_M}
\end{equation}
\begin{equation}
\varphi_{s\tau,s\tau'}=s\Theta_{\textrm{inter}}-(1-\tau)\frac{\pi}{2}\,,\label{eq:Intervalley_Hollow_varphi}
\end{equation}
with
\begin{equation}
\mathcal{M}_{\textrm{inter}}=2\prod_{\varsigma=\pm1}\frac{\Delta+\varsigma\Delta_{\textrm{so}}}{|d_{\varsigma}|}\,,\label{eq:Def_Minter}
\end{equation}
 
\begin{eqnarray}
\Theta_{\textrm{inter}}=-\sum_{\varsigma=\pm1}\varsigma\left[\arctan\left(\frac{\Im d_{\varsigma}}{\Re d_{\varsigma}}\right)+\pi H(-\Re d_{\varsigma})\right]\,,\label{eq:Def_Theta_inter}
\end{eqnarray}
and $H(.)$ is the Heaviside step function. Generally, both $\mathcal{M}_{\textrm{inter}}\neq0$
and $\Theta_{\textrm{inter}}\neq0$ and hence intervalley scattering
induced by a hollow-position adatom is skewed. The case of spin-preserving
intravalley scattering is similar:
\begin{equation}
\mathcal{M}_{s\tau,s\tau}=\tau\mathcal{M}_{\textrm{intra}}\,,
\end{equation}
\begin{equation}
\varphi_{s\tau,s\tau}=s\Theta_{\textrm{intra}}-(1-\tau)\frac{\pi}{2}\,,
\end{equation}
with 
\begin{equation}
\mathcal{M}_{\textrm{intra}}=2\prod_{\varsigma=\pm1}\frac{|\chi_{\varsigma}-p_{\varsigma}g|}{|d_{\varsigma}|}\,,
\end{equation}
and
\begin{eqnarray}
\Theta_{\textrm{intra}}-\Theta_{\textrm{inter}} & = & \sum_{\varsigma=\pm1}\varsigma\left[\arctan\left(\frac{\Im g}{\Re g-\frac{\chi_{\varsigma}}{p_{\varsigma}}}\right)\right.\nonumber \\
 &  & \left.+\pi H\left(\Re g-\frac{\chi_{\varsigma}}{p_{\varsigma}}\right)\right]\,.\label{eq:Def_Theta_intra}
\end{eqnarray}

\begin{figure}
\begin{centering}
\includegraphics[width=1\columnwidth]{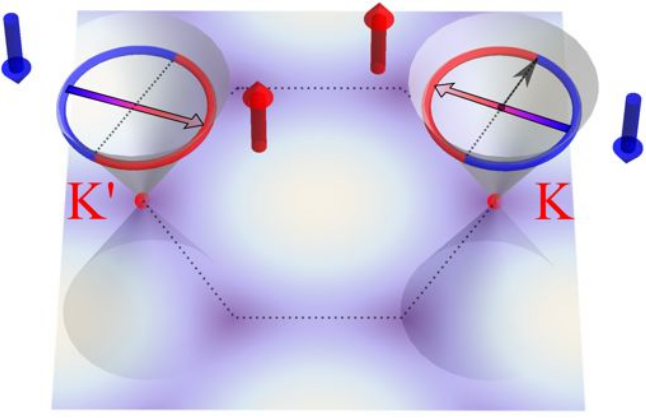}
\par\end{centering}

\caption{\label{fig:BrillouinScatt}Schematic representation of skew-scattering
induced by adatoms in hollow position. The black arrow represents
the momentum $\mathbf{k}_{\textrm{in}}$ of an incoming quasi-particle
in $K$ valley. Blue (red) half circles correspond to the region of
the Fermi line where the outgoing momentum $\mathbf{k}_{\textrm{out}}$
is most likely to be after a scattering event, if the incoming charge
carrier has spin down (spin up). Resulting pure spin currents are
depicted as blue and red planar arrows. Spin currents in $K$ and
$K'$ valleys associated with intra- and intervalley scattering tend
to oppose each other.}
\end{figure}
These results trivially lead to a zero transverse charge micro-current
\begin{equation}
J_{C}^{\perp}\equiv\sum_{s,\tau,\tau'}J_{s\tau,s\tau'}^{\perp}=0,
\end{equation}
but to a generally non-zero transverse spin micro-current
\begin{eqnarray}
J_{S}^{\perp} & \equiv & \sum_{s,\tau,\tau'}sJ_{s\tau,s\tau'}^{\perp}\nonumber \\
 & = & \frac{ks_{E}v_{F}}{2}\mathcal{M}_{\textrm{inter}}\left(\sin\Theta_{\textrm{inter}}-\mathcal{F}\sin\Theta_{\textrm{intra}}\right)\,,\label{eq:Transv_Spin_Curr}
\end{eqnarray}
where 
\begin{equation}
\mathcal{F}=\frac{\mathcal{M}_{\textrm{intra}}}{\mathcal{M}_{\textrm{inter}}}\,.
\end{equation}
The key parameters controlling the magnitude of the transverse spin
probability current $J_{S}^{\perp}$ are thus the phase difference
$\vartheta=\Theta_{\textrm{intra}}-\Theta_{\textrm{inter}}$ and the
$\mathcal{F}$ factor. They depend on the hopping energies characterizing
the graphene--adatom hybridization, and are thus expected to depend
strongly on the valence orbital type. Since \emph{s}-orbital adatoms
lack $J=\pm3/2,\pm5/2$ total angular momentum states, necessary to
couple hexagonal states of angular momentum $m=\pm2$, they induce
zero $\nu_{2}^{\pm},\Lambda_{R}$ {[}cf. Eq.~(\ref{eq:Hhollow_simple}){]},
directly leading to $p_{\pm}=0$. Consequently, $\vartheta=0$, $\mathcal{F}=1$
and thus $J_{S}^{\perp}=0$ for \emph{s}-orbital adatoms. Interestingly,
\emph{p}-orbital adatoms are a limiting case: they host exactly two
orbital states $J=\pm3/2$ and no $J=\pm5/2$ states. Therefore, $\Omega_{\pm2}^{\dagger}|0\rangle\longrightarrow\Omega_{\pm2}^{\dagger}|0\rangle$
transitions require double spin-flips, leading to small $\nu_{2}^{\pm}\propto w_{2}^{2}$
couplings, where $w_{2}$ is the hopping energy between graphene's
$\Omega_{\pm2,\downarrow/\uparrow}^{\dagger}|0\rangle$ state and
adatom's \emph{p} orbital with angular momentum $\pm1$ and spin-$\uparrow,\downarrow$;
see the Appendix. Generally, \emph{p}-orbital adatoms are thus expected
to yield negligible $\vartheta$, $\mathcal{F}\approx1$ and hence
small $J_{S}^{\perp}$. The case of \emph{d}- and \emph{f}-orbital
adatoms are noticeably different as they offer spin-preserving channels
for $\Omega_{\pm2}^{\dagger}|0\rangle\longrightarrow\Omega_{\pm2}^{\dagger}|0\rangle$
transitions, generally leading to appreciable $\nu_{2}^{\pm}$ couplings.
In the low-energy limit $|g|\ll\chi_{\pm}/p_{\pm}$, the $\mathcal{F}$
factor reads 
\begin{equation}
\mathcal{F}\approx\left(\frac{\nu_{1}^{+}+\nu_{2}^{+}}{\nu_{1}^{+}-\nu_{2}^{+}}\right)^{2}\neq1\,,
\end{equation}
where couplings corresponding to spin-dependent processes are neglected.
Clearly, this opens up the possibility of having large $J_{S}^{\perp}$
provided $\Theta_{\textrm{inter}}$ and/or $\vartheta$ are finite
(non-zero), leading to SHE.\cite{SHE,Jungwirth12} $\vartheta$ vanishes
in the vicinity of the Dirac point, reflecting the fact that intra-
and intervalley scattering mechanisms tend to yield transverse spin
currents of opposite signs, as depicted in Fig.~\ref{fig:BrillouinScatt}.
However, $\vartheta$ can become significant under certain conditions.
A natural question is thus whether $\vartheta$ can become large close
to the Dirac point for some physically meaningful values of $p_{\pm}$
and $\chi_{\pm}$. Typically, the function $\vartheta$ peaks when
$\Re g$ lies between $x_{+}=\frac{\chi_{+}}{p_{+}}$ and $x_{-}=\frac{\chi_{-}}{p_{-}}$.
Conditions $\Re g=x_{\pm}$ are fulfilled for Fermi levels $E_{X=x_{\pm}}$
verifying 
\begin{equation}
\frac{E_{X}}{E_{\textrm{c}}}\ln\left|\frac{E_{X}}{E_{\textrm{c}}}\right|=\frac{2\pi E_{\hexagon}^{2}X}{E_{\textrm{c}}}\,,\label{eq:Dirac_Point_Condition}
\end{equation}
where constants $\hbar v_{F}$ and $A_{\hexagon}$ have been restored,
and $E_{\hexagon}=\hbar v_{F}/\sqrt{A_{\hexagon}}$. Peak values of
$\vartheta$ are thus attained close to the Dirac point provided that
$2\pi E_{\hexagon}^{2}|X|\ll E_{c}$, in which case we find
\begin{equation}
\frac{E_{X}}{E_{c}}=\mathcal{L}\left(\frac{2\pi E_{\hexagon}^{2}X}{E_{\textrm{c}}}\right)\,.
\end{equation}
Here, $\mathcal{L}$ is a function defined by means of the lower branch
of the Lambert W-function, $\mathcal{W}_{-1}$:
\begin{eqnarray}
\mathcal{L}(y) & = & \frac{y}{\mathcal{W}_{-1}(-|y|)}\nonumber \\
 & \approx & \frac{y}{\ln|y|}\left(1+\frac{\ln|\ln|y||}{\ln|y|}+\frac{\ln^{2}|\ln|y||}{\ln^{2}|y|}-\frac{\ln|\ln|y||}{\ln^{2}|y|}\right)\,.\nonumber \\
\label{eq:Lambert}
\end{eqnarray}
We now determine under which conditions, $2\pi E_{\hexagon}^{2}|X|\ll E_{c}$
is verified. Parameters $p_{\pm}$ and $\chi_{\pm}$ can be expressed
in terms of the adatom's energy levels as well as tight-binding parameters
connecting hexagonal states to the adatoms valence orbital. If the
Fermi energy lies far away from the adatom's valence orbital energy
levels
\begin{equation}
\chi_{\pm}\approx-9\left(\frac{\Upsilon_{1}^{2}}{E_{1\pm1/2}^{+}}+\frac{\Upsilon_{2}^{2}}{E_{2\mp1/2}^{-}}\right)\,,\label{eq:chi}
\end{equation}
\begin{equation}
p_{\pm}\approx324\frac{\Upsilon_{1}^{2}\Upsilon_{2}^{2}}{E_{1\pm1/2}^{-}E_{2\mp1/2}^{+}}\,,\label{eq:p}
\end{equation}
where $E_{J}^{\pm}$ are energy levels that valence orbitals would
have in absence of intra-atomic spin-flip, $\Upsilon_{m}=u_{m}+\nu_{m}$
and $u_{m}$ and $\nu_{m}$ are hopping integrals connecting hexagonal
states and adatom's orbitals of same angular momentum $m$; see Eq.~(\ref{eq:TrueHgr-adHollow-1})
and text therein for definitions of the hopping integrals. We thus
have 
\begin{equation}
\frac{2\pi E_{\hexagon}^{2}x_{\pm}}{E_{\textrm{c}}}\approx-\frac{\pi E_{\hexagon}^{2}}{18E_{\textrm{c}}}\left(\frac{E_{1\pm1/2}^{+}}{\Upsilon_{1}^{2}}+\frac{E_{2\mp1/2}^{-}}{\Upsilon_{2}^{2}}\right)\,,
\end{equation}
which are small provided $E_{\textrm{c}}\Upsilon_{1,2}^{2}/E_{\hexagon}^{2}$
are significantly larger than the adatom's valence orbital energy
levels. 

In addition, $\Theta_{\textrm{inter}}$ exhibits resonances of its
own, which typically occur in energy windows where the real part of
$d_{\pm}$ is small. The real part of $d_{\pm}$ vanishes close to
the Dirac point at energies $E_{D=d_{\pm}}$, which relate to $X=x_{\pm}$
and $C=c_{\pm}=\frac{p_{\pm}}{\chi_{\pm}^{2}}$ according to 
\begin{equation}
\frac{E_{D}}{E_{\textrm{c}}}\approx\mathcal{L}\left(\frac{2\pi E_{\hexagon}^{2}X}{E_{\textrm{c}}}[1-\sqrt{1-C}]\right)\,.
\end{equation}
The function $\Theta_{\textrm{inter}}$ exhibits resonances close
to the Dirac point provided that $2\pi E_{\hexagon}^{2}|X|\ll E_{\textrm{c}}$
or $C\ll1$. The former condition is valid whenever both $E_{\textrm{c}}\Upsilon_{1}/E_{\hexagon}^{2}$
and $E_{\textrm{c}}\Upsilon_{2}/E_{\hexagon}^{2}$ are large compared
to adatom energy levels, whereas, the latter condition is fulfilled
if $\Upsilon_{1}\gg\Upsilon_{2}$ or $\Upsilon_{2}\gg\Upsilon_{1}$.
More precisely, $c_{\pm}\propto(\Upsilon_{2}/\Upsilon_{1})^{2}$ whenever
$\Upsilon_{1}\gg\Upsilon_{2}$ and $c_{\pm}\propto(\Upsilon_{1}/\Upsilon_{2})^{2}$
in the opposite limit. 

\begin{figure}
\begin{centering}
\includegraphics[width=0.8\columnwidth]{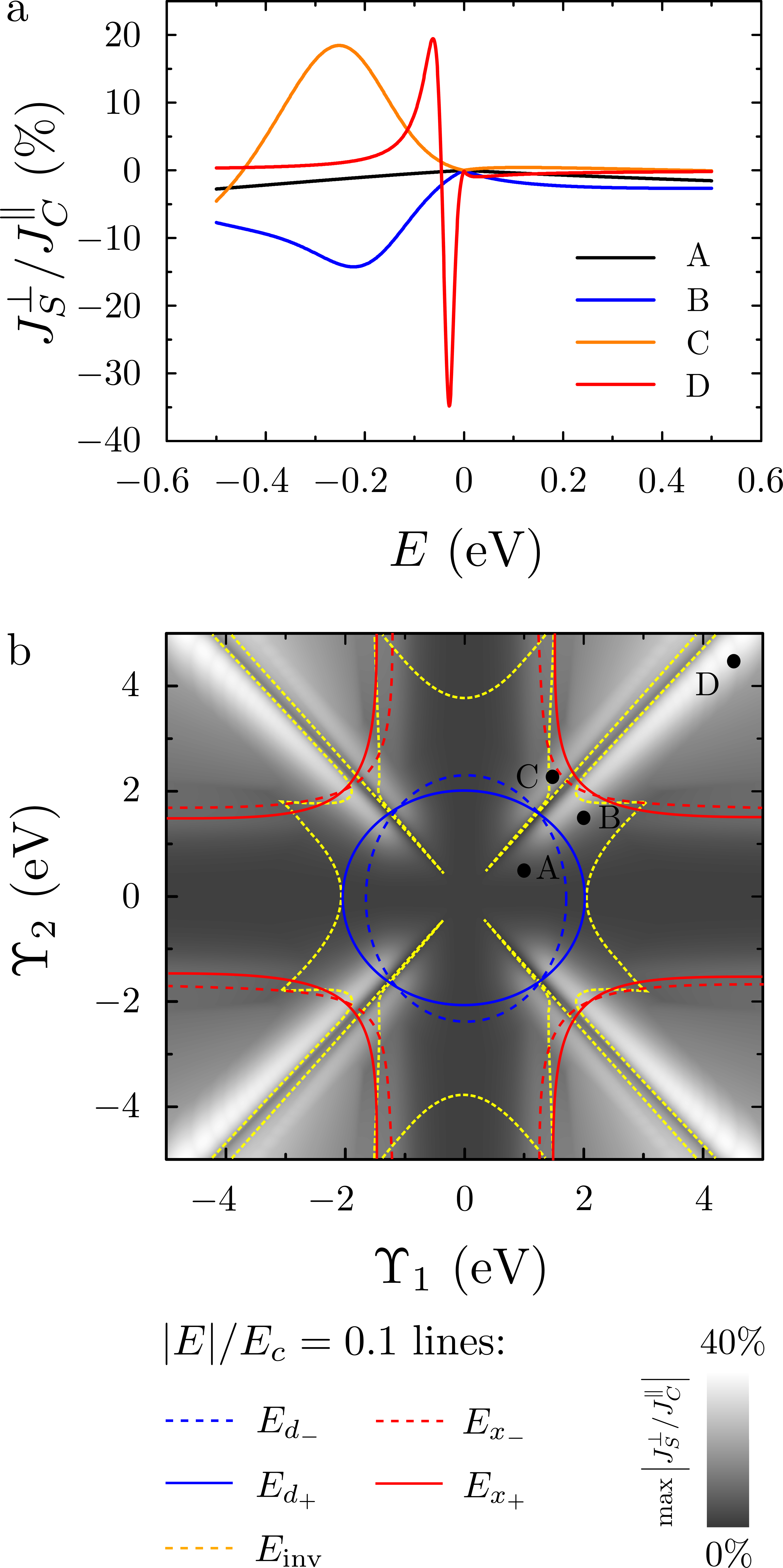}
\par\end{centering}

\caption{\label{fig:SHE}Resonant features of spin transverse currents generated
by hollow-position physisorbed adatoms on graphene. (a) $J_{S}^{\perp}/J_{C}^{\parallel}$
(in \%) against Fermi energy (in eV), for adatoms with $E_{1/2}^{\pm}=-1$~eV,
$E_{3/2}^{\pm}=-1.5$~eV, and $E_{5/2}^{\pm}=-2$~eV and different
$(\Upsilon_{1},\Upsilon_{2})$, corresponding to points A, B, C and
D shown in the lower panel. (b) Maximum of $|J_{S}^{\perp}/J_{C}^{\parallel}|$
for $|E_{F}|\leq0.5$~eV, against $\Upsilon_{1}$ and $\Upsilon_{2}$.
$|E|/E_{\textrm{c}}=0.1$ lines are shown for $E=E_{x_{\pm},d_{\pm}},E_{\textrm{inv}}$.
Each line partitions the $(\Upsilon_{1},\Upsilon_{2})$ space into
two parts: a region containing the origin $\Upsilon_{1}=\Upsilon_{2}=0$,
characterized by high-energy resonances, $|E|/E_{c}>0.1$, and a region
away from the origin, characterized by low-energy resonances, $|E|/E_{c}<0.1$. }
\end{figure}

Figure~\ref{fig:SHE}(a) shows $J_{S}^{\perp}$ as a fraction of
the charge current 
\begin{equation}
J_{C}^{\parallel}=s_{E}v_{F}\sum_{s,\tau,s',\tau'}\Sigma_{s\tau,s'\tau'}^{\parallel}
\end{equation}
against Fermi energy $E_{F}$, for realistic values of hopping integrals
and atomic energy levels. While the adatom energy levels are kept
fixed, $J_{S}^{\perp}/J_{C}^{\parallel}$ is plotted for different
pairs of couplings $(\Upsilon_{1},\Upsilon_{2})$, corresponding to
points $A,B,C$ and $D$ shown in Fig.~\ref{fig:SHE}(b). Figure~\ref{fig:SHE}(a)
illustrates the strong dependence of $J_{S}^{\perp}/J_{C}^{\parallel}$
on couplings between the adatom valence orbitals and graphene hexagonal
states. At point $A$, the transverse spin micro-current is negligible
compared to $J_{C}^{\parallel}$, whereas points $B$ and $C$ yield
transverse spin currents as large as $\sim20\%$ of the total outgoing
current at resonance. In situation $D$, $J_{S}^{\perp}/J_{C}^{\parallel}$
exhibits giant peak-values of up to $0.4$ in magnitude. Figure~\ref{fig:SHE}(b)
connects the existence of peaks in $J_{S}^{\perp}/J_{C}^{\parallel}$
for particular $(\Upsilon_{1},\Upsilon_{2})$ points to previously
discussed resonant energies $E_{x_{\pm},d_{\pm}}$.\textcolor{black}{{}
It highlights that $E_{x_{\pm},d_{\pm}}$ and resulting peaks in transverse
spin currents exist at low energy for sufficiently large $|\Upsilon_{1}|$
or $|\Upsilon_{2}|$. However, the peak values of $J_{S}^{\perp}/J_{C}^{\parallel}$
only become significant when both $|\Upsilon_{1}|$ and $|\Upsilon_{2}|$
exceed $\sim E_{\hexagon}(E_{J}^{\pm}/E_{\textrm{c}})^{1/2}$. }

While transverse spin currents, arising from skew-scattering of graphene
Dirac fermions with a hollow-position adatom, can exhibit large resonances,
Eq.~(\ref{eq:Transv_Spin_Curr}) suggests that $J_{S}^{\perp}$ possesses
another interesting property. Since $J_{S}^{\perp}$ results from
competing transverse spin currents originating from intra- and intervalley
scattering, one expects $J_{S}^{\perp}$ to change sign for particular
Fermi energies, such that
\begin{equation}
\tan\Theta_{\textrm{inter}}=\frac{\mathcal{F\sin\vartheta}}{1-\mathcal{F}\cos\vartheta}\,.\label{eq:Spin_Curr_Inversion}
\end{equation}
The existence of such levels close to the Dirac point would open up
interesting technological prospects, as a field effect would allow
to reverse spin-current flows. \textcolor{black}{Intriguingly, Eq.~(\ref{eq:Spin_Curr_Inversion})
admits low-energy solutions for sufficiently large $|\Upsilon_{1}|$
or $|\Upsilon_{2}|$.} For the sake of clarity, points $(\Upsilon_{1},\Upsilon_{2})$
such that the solution $E_{\textrm{inv}}$ of minimum magnitude is
equal to a tenth of graphene half-bandwidth are shown in Fig.~\ref{fig:SHE}(b),
as a yellow dashed line. Energies $E_{\textrm{inv}}$ closer to the
Dirac point are obtained away from the origin, beyond the yellow curve.
This is illustrated by point $D$, whose corresponding $J_{S}^{\perp}/J_{C}^{\parallel}$
against the $E_{F}$ curve exhibits a sharp inversion in transverse
spin-current flow around $E_{F}\approx-50$ meV; see Fig.~\ref{fig:SHE}(a).

Finally, let us highlight further the specificities of the above-discussed
spin currents. Although spin Hall related phenomena of intrinsic\cite{Abanin-Geim}
and extrinsic\cite{Jaya} types have already been observed in graphene,
they differ drastically from the SHE discussed in our work. While
in Ref.~\onlinecite{Abanin-Geim}, the SHE necessitates a strong
magnetic field and relies on Zeeman splitting at the Dirac point,\cite{Abanin}
correlating spin $\uparrow,\downarrow$ and charge $\mp e$, SHE observed
in Ref.~\onlinecite{Jaya} is induced by the deformation of a graphene
lattice due to the presence of $sp^{3}$ bonds.\cite{AHCN-SOC09}
Our theory describes SHE arising from hybridization of graphene with
adatoms in the hollow position, and predicts the appearance of large
spin currents around resonant energies $E_{x_{\pm}}$ and $E_{d_{\pm}}$
for \emph{d}- or \emph{f}- orbital adatoms. The nature of these resonances
is graphene-specific. Scattering events with $\delta$-impurities
generally lead to outgoing waves that are linear combinations of Dirac
spinors. The corresponding coefficients are of the form $\varrho_{1}+\varrho_{2}e^{-i\theta}$
as shown by Eq.~(\ref{eq:c}). For scalar potentials, $\varrho_{1}=\varrho_{2}$,
leading to the well-known absence of back-scattering.\cite{Scattering_Theory_SLG}
However, for ``tensor potentials,'' such as the Kane-Mele term $V_{\textrm{so}}s_{z}\tau_{z}\sigma_{z}\delta(\mathbf{r})$,
the $T$~matrix is sublattice-dependent and $\varrho_{1}$ and $\varrho_{2}$
can thus acquire a phase difference, which gives rise to gate-tunable
$\Theta$\textsubscript{intra} and $\Theta$\textsubscript{inter}
{[}through $g(E)${]}, enabling the discussed resonances. The SHE
arising from hollow-position adatoms on graphene is thus significantly
different from extrinsic SHE phenomena typically studied in parabolic-band
2D electron gases, and the above-discussed resonances are unrelated
to previously-observed enhancement of skew scattering due to the orbital-dependent
Kondo effect\cite{Gu-SHE} or a large SOC energy band.\cite{Molenkamp}
Our calculations show that the efficiency of isolated adatoms to generate
transverse spin currents, i.e., $J_{S}^{\perp}/J_{C}^{\parallel}=\mathcal{O}(0.1)$,
is comparable to that recently predicted for large SOC-active clusters.\cite{SHE_G_14}
Last but not least, an interesting feature of the resonant regime
illustrated by Fig.~\ref{fig:SHE} is the possibility to change the
sign of $J_{S}^{\perp}$ upon tuning the Fermi level around specific
``inversion energies'' $E_{\textrm{inv}}$. A direct consequence
is the ability to convert a charge current into a large transverse
spin current in a certain energy range, and to reverse its flow by
tuning the gate voltage around critical values, which could find applications
in spin-based logic devices.

\subsection*{II.c. Scattering with top-position adatoms}

We now deal with the scattering mechanisms induced by an adatom in
the top position. We start with adatoms on top of an $A$-sublattice
carbon atom. Using Table~\ref{tab:EffectiveMass}, the corresponding
$T$~matrix in basis (\ref{eq:MagicBasis}) reads
\begin{equation}
T_{\textrm{top},A}=\left(\begin{array}{cc}
T_{\textrm{top},A}^{\uparrow\uparrow} & T_{\textrm{top},A}^{\downarrow\uparrow}\\
T_{\textrm{top},A}^{\uparrow\downarrow} & T_{\textrm{top},A}^{\downarrow\downarrow}
\end{array}\right)\,,
\end{equation}
with $4\times4$ blocks:
\begin{equation}
T_{\textrm{top},A}^{ss}=\left(\begin{array}{cccc}
a & 0 & 0 & a\\
0 & b_{s} & 0 & 0\\
0 & 0 & b'_{s} & 0\\
a & 0 & 0 & a
\end{array}\right)\,,\label{eq:Ttop1}
\end{equation}
where $s=\uparrow,\downarrow$ and 
\begin{equation}
T_{\textrm{top},A}^{\uparrow\downarrow}=-[T_{\textrm{top},A}^{\downarrow\uparrow}]^{\textrm{t}}=\left(\begin{array}{cccc}
0 & 0 & -t & 0\\
t & 0 & 0 & t\\
0 & 0 & 0 & 0\\
0 & 0 & -t & 0
\end{array}\right)\,.\label{eq:Ttop2}
\end{equation}
The $T$ matrix elements in Eqs.~(\ref{eq:Ttop1}) and (\ref{eq:Ttop2})
verify the following identities:
\begin{eqnarray}
t & = & \frac{2i\Lambda_{\textrm{so}}}{1-Ug+2wg^{2}}\,,\label{eq:top_element_1}
\end{eqnarray}
\begin{eqnarray}
a & = & \frac{V_{0}-wg}{1-Ug+2wg^{2}}\,,\label{eq:top_element_4}
\end{eqnarray}
\begin{equation}
b_{\uparrow}=b'_{\downarrow}=\frac{v_{0}+\lambda_{\textrm{so}}}{1-g(v_{0}+\lambda_{\textrm{so}})}\,,\label{eq:top_element_5}
\end{equation}
\begin{equation}
b_{\downarrow}=b'_{\uparrow}=\frac{v_{0}-\lambda_{\textrm{so}}-2wg}{1-Ug+2wg^{2}}\,,\label{eq:top_element_6}
\end{equation}
where we set $U=2V_{0}+v_{0}-\lambda_{\textrm{so}}$ and $w=(v_{0}-\lambda_{\textrm{so}})V_{0}-4\Lambda_{\textrm{so}}^{2}$.
The $T$~matrix for a $B$-sublattice adatom is easily obtained from
$T_{\textrm{top},A}$ by reflection $\mathcal{R}_{y}$, i.e.,
\begin{equation}
T_{\textrm{top},B}=\mathcal{U}_{y}T_{\textrm{top},A}\mathcal{U}_{y}\,.\label{eq:TtopA to TtopB}
\end{equation}
Matrix elements $t_{s\tau,s'\tau'}^{\sigma,\sigma'}$ and $\widetilde{t}_{s\tau,s'\tau'}^{\,\,\sigma,\sigma'}$
of $T_{\textrm{top},A}$ and $T_{\textrm{top},B}$, respectively,
associated with $s\tau\sigma\rightarrow s'\tau'\sigma'$ transitions,
are thus related by 
\begin{equation}
\widetilde{t}{}_{s\tau,s'\tau'}^{\,\,\sigma,\sigma'}=ss't_{-s\tau,-s'\tau'}^{\,\overline{\sigma},\overline{\sigma}'}\,,\label{eq:TtopB_elements}
\end{equation}
with $\overline{A}=B$ and $\overline{B}=A$. We next describe the
possible scattering mechanisms induced by an adatom on top of a $\circ$-sublattice
site, $\circ=A,B$, by calculating corresponding cross-sections $\sigma_{s\tau,s'\tau'}^{\circ}$.
Since $T_{\textrm{top},A}$ transforms into $T_{\textrm{top},B}$
under $\mathcal{R}_{y}$, the following relation holds:
\begin{equation}
\sigma_{s\tau,s'\tau'}^{B}(\theta)=\sigma_{-s\tau,-s'\tau'}^{A}(-\theta)\,,\label{eq:Cross-sections_TopA_Top_B}
\end{equation}
so that we can focus on computing $\sigma_{s\tau,s'\tau'}^{A}(\theta)$
only.

From this perspective, we first describe the scattering mechanisms
that do not conserve spin and valley. Equation~(\ref{eq:Ttop1})
directly implies that intravalley spin-flip and spin-preserving intervalley
scattering induced by top-position adatoms are isotropic mechanisms,
as
\begin{equation}
\sigma_{s\tau,-s\tau}^{A}=\frac{k|t|^{2}}{8\pi},\label{eq:top_spin-flip}
\end{equation}
 and
\begin{equation}
\sigma_{s\tau,s-\tau}^{A}=\frac{k|a|^{2}}{8\pi}\,.\label{eq:top_inter-valley}
\end{equation}
Unlike adatoms in the hollow position, top-position adatoms induce
intervalley spin-flip scattering processes {[}see Eq.~(\ref{eq:d_pm})
and text therein{]}. In particular, for adatoms on the $A$-sublattice,
corresponding differential cross-sections, for $s\neq s'$ and $\tau\neq\tau'$,
are finite:
\begin{equation}
\sigma_{s\tau,s'\tau'}^{A}(\theta)=\frac{k|t|^{2}}{2\pi}\cos^{2}\left(\frac{\theta}{2}\right)\delta_{s+\tau}\,.\label{eq:sig_top_A_intervalley_flip}
\end{equation}
However, since (\ref{eq:sig_top_A_intervalley_flip}) is an even function
of $\theta$, scattering mechanisms originating from top-position
adatoms yield zero transverse currents, i.e., $J_{s\tau,s'\tau'}^{A\perp}=0$,
whenever $s\neq s'$ or $\tau\neq\tau'$.\textcolor{black}{{} These
general considerations are consistent with first-principles calculations
showing that spin-relaxation rates in graphene are very sensitive
to the adsorption site.\cite{Federov2013} }

Next, we study spin-preserving intravalley scattering. Irrespective
of the valley $\tau$ and spin $s$, $T$-matrix elements associated
with the $A$-sublattice $t_{s\tau,s\tau}^{A,A}$ are equal. This
contrasts with $t_{s\tau,s\tau}^{B,B}$ elements, which generally
verify 
\begin{equation}
t_{\uparrow K,\uparrow K}^{B,B}=t_{\downarrow K',\downarrow K'}^{B,B}\neq t_{\uparrow K',\uparrow K'}^{B,B}=t_{\downarrow K,\downarrow K}^{B,B}\,.
\end{equation}
As a result, spin-preserving intravalley-scattering cross sections
for $\uparrow K$ and $\downarrow K'$ charge carriers differ from
those for $\downarrow K$ and $\uparrow K'$ quasi-particles, and
$\sigma_{s\tau,s\tau}(\theta)$ is determined by the conserved quantity
$s+\tau$. We start by considering the $s+\tau=\pm2$ case. We define
$\mathcal{M}_{2}=\mathcal{M}_{\uparrow K,\uparrow K}^{A}=\mathcal{M}_{\downarrow K',\downarrow K'}^{A}$
and $\varphi_{2}=\mathcal{\varphi}_{\uparrow K,\uparrow K}^{A}=\mathcal{\varphi}_{\downarrow K',\downarrow K'}^{A}$,
which verify 
\begin{equation}
\mathcal{M}_{2}=\frac{2|v_{0}+\lambda_{\textrm{so}}|.|V_{0}-wg|}{|1-Ug+2wg^{2}|.|1-g(v_{0}+\lambda_{\textrm{so}})|}\,,\label{eq:M2}
\end{equation}
and
\begin{eqnarray}
\varphi_{2} & = & \arctan\left(\frac{U\Im g-2w\Im(g^{2})}{1-U\Re g+2w\Re(g^{2})}\right)\nonumber \\
 & + & \arctan\left(\frac{(v_{0}+\lambda_{\textrm{so}})\Im g}{(v_{0}+\lambda_{\textrm{so}})\Re g-1}\right)\nonumber \\
 & + & \arctan\left(\frac{w\Im g}{w\Re g-V_{0}}\right)\nonumber \\
 & + & \pi H\left(w\Re g-V_{0}\right)+\pi H\left[(v_{0}+\lambda_{\textrm{so}})\Re g-1\right]\nonumber \\
 & - & \pi H\left[U\Re g-2w\Re(g^{2})-1\right]\,.\label{eq:PureMomPhase_top}
\end{eqnarray}
The case of Dirac fermions for which $s+\tau=0$ is markedly different.
Denoting $\mathcal{M}_{0}=\mathcal{M}_{\downarrow K,\downarrow K}^{A}=\mathcal{M}_{\uparrow K',\uparrow K'}^{A}$
and $\varphi_{0}=\mathcal{\varphi}_{\downarrow K,\downarrow K}^{A}=\mathcal{\varphi}_{\uparrow K',\uparrow K'}^{A}$,
we obtain 
\begin{equation}
\mathcal{M}_{0}=\frac{2|V_{0}-wg|.|v_{0}-\lambda_{\textrm{so}}-2wg|}{|1-Ug+2wg^{2}|^{2}}\,,\label{eq:M0}
\end{equation}
and
\begin{eqnarray}
\varphi_{0} & = & \arctan\left(\frac{w\Im g}{V_{0}-w\Re g}\right)\nonumber \\
 & - & \arctan\left(\frac{2w\Im g}{v_{0}-\lambda_{\textrm{so}}-2w\Re g}\right)\nonumber \\
 & + & \pi H\left(2w\Re g+\lambda_{\textrm{so}}-v_{0}\right)-\pi H\left(w\Re g-V_{0}\right)\,.\nonumber \\
\label{eq:PureMomPhase_0_top}
\end{eqnarray}
Crucially, currents $J_{\uparrow K,\uparrow K}^{A\perp}+J_{\uparrow K',\uparrow K'}^{A\perp}$
and $J_{\downarrow K,\downarrow K}^{A\perp}+J_{\downarrow K',\downarrow K'}^{A\perp}$
are equal, so that spin-preserving intravalley scattering does not
give rise to any transverse spin current. The same holds true for
spin-preserving intravalley scattering induced by an adatom on the
$B$-sublattice, but due to relation (\ref{eq:Cross-sections_TopA_Top_B}),
\begin{equation}
J_{s\tau,s\tau}^{B\perp}=-J_{-s\tau,-s\tau}^{A\perp}\,.\label{eq:Curr_A-Curr_B}
\end{equation}
However, the transverse charge current arising from scattering with
a single top-position adatom on sublattice $\sigma=A,B$
\begin{equation}
J_{C}^{\sigma\perp}=\sum_{s,\tau}J_{s\tau,s\tau}^{\sigma\perp}\,,\label{eq:General_Transverse_Curr}
\end{equation}
is generally non-zero, and Eq.~(\ref{eq:Curr_A-Curr_B}) directly
implies $J_{Q}^{A\perp}=-J_{Q}^{B\perp}$. Macroscopically, top-position
adatoms thus give rise to the CHE, provided that the populations of
adatoms on the $A$ and $B$ sublattices differ by type or number.
Let us now study transverse charge currents $J_{C}^{\sigma\perp}$
in more detail. Using Eqs.~(\ref{eq:General_Transverse_Curr}), (\ref{eq:long_micro_current}),
and (\ref{eq:General_cross-section}), we obtain 
\begin{equation}
J_{C}^{\sigma\perp}=\mp\frac{ks_{E}v_{F}}{4}(\mathcal{M}_{0}\sin\varphi_{0}+\mathcal{M}_{2}\sin\varphi_{2})\,,
\end{equation}
with $-$~$(+)$ for adatoms on the $A$~$(B)$ sublattice. It is
interesting to note that the magnitude of $J_{C}^{\sigma\perp}$ is
modulated by $\mathcal{M}_{0}$ and $\mathcal{M}_{2}$, which are
proportional to $|v_{0}-\lambda_{\textrm{so}}-2wg|$ and $|v_{0}+\lambda_{\textrm{so}}|$
respectively. While the dependence on $\lambda_{\textrm{so}}$ is
expected---as spin--orbit interaction is a well-known cause of anomalous
Hall effect\cite{AHE}---the dependence on $v_{0}$, a scalar potential
acting on graphene triangular states $\Gamma_{m,s}^{\dagger}|0\rangle$,
is more surprising. However, this $v_{0}$-dependence has a trivial
geometrical explanation: the $v_{0}\pi_{B,A}$ term of Hamiltonian
$\mathcal{H}_{\textrm{top}}^{A,B}$ is the continuum-theory counterpart
of the trigonal potential which affects the three graphene $p_{z}$
orbitals neighboring the adsorption site, and trigonal potentials
clearly scatter charges anisotropically.

We next describe the energy-dependence of $J_{C}^{\sigma\perp}$.
In neutral graphene, phases $\varphi_{0}$ and $\varphi_{2}$ are
null. However, they exhibit large resonances at finite Fermi energies,
such that
\begin{equation}
\Re g=1/\omega_{i=1,2,3}\,,\label{eq:top_resonance_type_I}
\end{equation}
where
\begin{equation}
\omega_{1}=v_{0}+\lambda_{\textrm{so}}\,,
\end{equation}
\begin{equation}
\omega_{2}=v_{0}-\lambda_{\textrm{so}}-\frac{4\Lambda_{\textrm{so}}^{2}}{V_{0}}\,,
\end{equation}
\begin{equation}
\omega_{3}=2V_{0}-\frac{8\Lambda_{\textrm{so}}^{2}}{v_{0}-\lambda_{\textrm{so}}}\,.
\end{equation}
Equation (\ref{eq:top_resonance_type_I}) has a low-energy root $|\mathcal{E}_{i}|\ll E_{c}$
provided that $|\omega_{i}|\gg2\pi E_{\hexagon}^{2}/E_{c}$, in which
case
\begin{equation}
\frac{\mathcal{E}_{i}}{E_{\textrm{c}}}=\mathcal{L}\left(\frac{2\pi E_{\hexagon}^{2}}{E_{\textrm{c}}\omega_{i}}\right)\,.
\end{equation}
Another resonance is reached whenever 
\begin{equation}
\Re(g^{2})-\frac{U}{2w}\Re g+\frac{1}{2w}=0\,.\label{eq:top_resonance_type_II}
\end{equation}
Equation (\ref{eq:top_resonance_type_II}) admits a low energy solution
$\mathcal{E}_{4}$ verifying 
\begin{equation}
\frac{\mathcal{E}_{4}}{E_{c}}\approx\mathcal{L}\left(\frac{2\pi E_{\hexagon}^{2}}{E_{c}U}\right)\,,\label{eq:Eres}
\end{equation}
provided $8|w|\ll U^{2}$ and $|U|\gg2\pi E_{\textrm{c}}$. 

We now write conditions for the existence of resonant energies close
to the Dirac point, in terms of tight-binding parameters connecting
central and triangular states to the top-position adatoms. We first
consider the marginal case of \emph{s}-orbital adatoms, which only
host states of total angular momentum $J=\pm1/2$. Couplings between
triangular states of angular momentum $m=\pm1$ are thus necessarily
mediated by double spin-flip through an available adatom orbital.
Using the Appendix notations as well as Eqs.~(\ref{eq:Stop_1p5})
and (\ref{eq:Stop_0p5}) within, this results in $v_{0}=-\lambda_{\textrm{so}}\approx-\frac{9}{2}\frac{l_{\textrm{so},1}^{2}}{E_{1/2}^{+}}$
and $\Lambda_{\textrm{so}}\approx-\frac{9}{2}\frac{l_{\textrm{so},1}\gamma}{E_{1/2}^{+}}$.
Therefore, $\omega_{1}=0$ and resonant energy $\mathcal{E}_{1}$
is infinite. In addition, the hopping integral $l_{\textrm{so},1}$
connecting triangular states $\Gamma_{\pm1,\downarrow/\uparrow}^{\dagger}|0\rangle$
to \emph{s} orbitals of opposite spin is expected to be small compared
to $E_{1/2}^{+}$ and graphene half-bandwidth $E_{\textrm{c}}$, leading
to $|\omega_{2,3}|\ll2\pi E_{\hexagon}^{2}/E_{\textrm{c}}$ and $|\mathcal{E}_{2,3}|\gg E_{\textrm{c}}$.
Resonant energies $\mathcal{E}_{1,2,3}$ are thus experimentally irrelevant.
However, $U\approx-2\gamma^{2}/E_{1/2}^{+}$ and $\mathcal{E}_{4}$
is the only resonance which can possibly be observed, provided $\gamma^{2}/|E_{1/2}^{+}|\gg\pi E_{\textrm{c}}$. 

For all other types of valence orbitals, i.e., \emph{p}, \emph{d},
and \emph{f}, the existence of spin-preserving channels coupling triangular
states of angular momentum $m=\pm1$ gives rise to enhanced $v_{0}$,
$\lambda_{\textrm{so}}$ and $\Lambda_{\textrm{so}}$ energy scales,
making resonant energies $\mathcal{E}_{i=1,2,3}$ accessible under
certain conditions. Equations~(\ref{eq:Stop_1p5}) and (\ref{eq:Stop_0p5})
directly lead to:

\begin{equation}
V_{0}\sim-\frac{\gamma^{2}}{E_{1/2}^{+}}\,,
\end{equation}
\begin{equation}
v_{0}\sim-\frac{9}{2}\theta_{1}^{2}\left(\frac{1}{E_{1/2}^{-}}+\frac{1}{E_{3/2}^{+}}\right)\,,
\end{equation}
\begin{equation}
\lambda_{\textrm{so}}\sim\frac{9}{2}\theta_{1}^{2}\left(\frac{1}{E_{1/2}^{-}}-\frac{1}{E_{3/2}^{+}}\right)\,,
\end{equation}
\begin{equation}
\Lambda_{\textrm{so}}\sim-\frac{9}{2}\frac{\gamma\theta_{1}\Lambda_{\textrm{so}}^{1}}{E_{1/2}^{-}E_{1/2}^{+}}\,.
\end{equation}
Therefore, Eq.~(\ref{eq:top_resonance_type_I}) holds for $i=1,2$
or $3$ if $\theta_{1}^{2}/|E_{3/2}^{+}|\gg E_{\hexagon}^{2}/E_{\textrm{c}}$,
$\theta_{1}^{2}/|E_{1/2}^{-}|\gg E_{\hexagon}^{2}/E_{\textrm{c}}$
or $\gamma^{2}/|E_{1/2}^{+}|\gg2\pi E_{\hexagon}^{2}/E_{\textrm{c}}$
respectively. Moreover, 
\begin{equation}
U\sim-\frac{2\gamma^{2}}{E_{1/2}^{+}}-\frac{9\theta_{1}^{2}}{E_{1/2}^{-}}\,,
\end{equation}
and assuming $\Lambda_{\textrm{so}}$ is much smaller than $V_{0}$,
$v_{0}$ and $\lambda_{\textrm{so}}$, we have
\begin{equation}
w\sim\frac{9\theta_{1}^{2}\gamma^{2}}{E_{1/2}^{-}E_{1/2}^{+}}\,,
\end{equation}
\begin{figure}
\begin{centering}
\includegraphics[width=0.8\columnwidth]{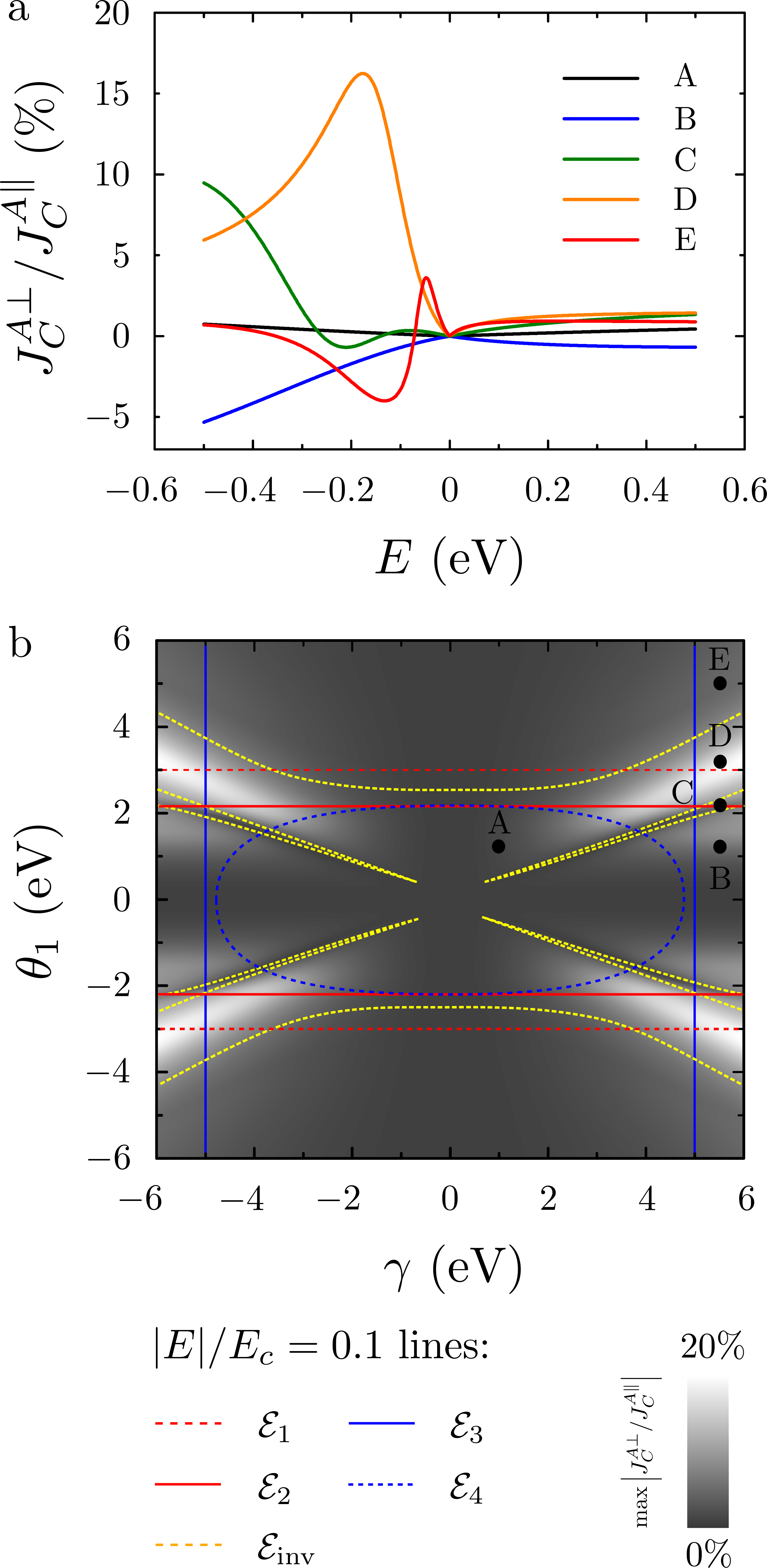}
\par\end{centering}

\caption{\label{fig:AHE}Resonant features of charge transverse currents generated
by top-position physisorbed adatoms on graphene. (a) $J_{C}^{A\perp}/J_{C}^{A\,\parallel}$
(in \%) against Fermi energy (in eV), for top-position adatoms with
fixed $E_{1/2}^{\pm}=-1$~eV and $E_{3/2}^{\pm}=-1.5$~eV and different
$(\gamma,\theta_{1})$, corresponding to points A, B, C, D and E shown
in lower panel. (b) Maximum of $|J_{C}^{A\perp}/J_{C}^{A\,\parallel}|$
for $|E_{F}|\leq0.5$ eV, against $\gamma$ and $\theta_{1}$. $|E|/E_{c}=0.1$
lines are shown for $E=\mathcal{E}_{1,2,3,4},\mathcal{E}_{\textrm{inv}}$.
Similarly to Fig.~\ref{fig:SHE}, each line partitions $(\gamma,\theta_{1})$-space
into regions, whose farthest from the origin corresponds to $|E|/E_{c}<0.1$. }
\end{figure}
so that Eq.~(\ref{eq:Eres}) is valid provided $|\gamma|\gg|\theta_{1}|$
or $|\gamma|\ll|\theta_{1}|$, and $2\gamma^{2}/|E_{1/2}^{+}|+9\theta_{1}^{2}/|E_{1/2}^{-}|\gg2\pi E_{\hexagon}^{2}/E_{c}$. 

Figure \ref{fig:AHE}(a) shows the transverse charge current $J_{C}^{A\perp}$
as a fraction of total outgoing current 
\begin{equation}
J_{C}^{A\,\parallel}=s_{E}v_{F}\sum_{s,\tau,s',\tau'}\Sigma_{s\tau,s'\tau'}^{A\,\parallel}\,,
\end{equation}
against Fermi energy $E_{F}$, for fixed values of atomic energy levels
$E_{1/2}^{\pm}$, and $E_{3/2}^{\pm}$, and various $(\gamma,\theta_{1})$
points, labeled as $A,B,C,D$ and $E$. While for small $\gamma$
and $\theta_{1}$ (situation $A$), the transverse charge current
is negligible compared to $J_{C}^{A\,\parallel}$, significant $J_{C}^{A\perp}$
currents are obtained for values of $\gamma$ and $\theta_{1}$ of
the order of few eV (points $B,C,D,E$ in Fig.~\ref{fig:AHE}(b)),
up to $20\%$. In addition, the transverse charge current can change
direction for some values of Fermi energy, as illustrated by curves
$C$ and $E$ in Fig.~\ref{fig:AHE}(a). Such ``inversion'' energies
can exist close to the Dirac point for finite values of $\gamma$
and $\theta_{1}$ only. Noting $\mathcal{E}_{\textrm{inv}}$ the inversion
energy closest to the Dirac point for a given $(\gamma,\theta_{1})$
couple, Fig.~\ref{fig:AHE}(b) shows $|\mathcal{E}_{\textrm{inv}}(\gamma,\theta_{1})|=E_{c}/10$
lines, which partition $(\gamma,\theta_{1})$-space into regions whose
farthest from the origin corresponds to $|\mathcal{E}_{\textrm{inv}}|/E_{c}<0.1$
. Clearly, $|\mathcal{E}_{\textrm{inv}}|/E_{c}<0.1$ domains overlap
with regions of large $J_{C}^{A\perp}/J_{C}^{A\,\parallel}$ magnitude,
making the existence of $\mathcal{E}_{\textrm{inv}}$ relevant for
applications. Similar to transverse spin currents arising from scattering
with hollow-position adatoms, we believe that the possibility of changing
the sign of $J_{Q}^{\sigma\perp}$ by field effect could be explored
for new functionalities in logic devices. At the scale of an entire
graphene device however, the observation of significant macroscopic
transverse charge currents due to scattering with a large ensemble
of top-position adatoms appears more challenging than the observation
of large SHE due to hollow-position adatoms, because of the necessity
of having an imbalance between $A$- and $B$-sublattice. Nevertheless,
it should be noted that sublattice ordering driven by RKKY-type interactions
below a critical temperature was predicted by several authors,\cite{Levitov10,Cheianov09,Cheianov10,Kopylov11}
so that the above-discussed CHE may in principle be observed in an
experiment.\textcolor{red}{{} }

\section{Concluding remarks}

In this work we have provided a rigorous derivation of effective graphene--adatom
Hamiltonians, taking into account intervalley terms neglected in previous
works. Our results describe the experimentally relevant scenario of
dilute physically adsorbed adatoms randomly distributed over the graphene
lattice. We have shown that both the position in the lattice and the
valence orbital type are critical to determine the action of a physisorbed
adatom on graphene's Dirac fermions. Our study of non-magnetic elements---valid
when the Fermi energy is detuned from the adatoms valence orbital
spectrum---established that while bridge-position adsorption does
not induce spin--orbit coupling, hollow- and top-position adatoms
can generate significant local spin--orbit interactions, in such a
way that spin and valley quantum numbers are strongly intertwined.
The low-energy continuum theories constructed for hollow- and top-position
species allowed us to obtain analytic expressions for $T$ matrices
and integrated cross sections and to derive the corresponding charge
carriers' spin-dependent scattering mechanisms. Interestingly, hollow-
and top-position spin--orbit active adatoms give rise to Hall effects
of drastically different nature: pure spin currents for the former
(SHE), and non-polarized charge currents for the latter (CHE). They
nonetheless have two key characteristics in common: they can be switched
on and off and their flow can be reversed by tuning the Fermi energy.
We anticipate that such properties will find technological applications
in the fields of spin- and charge-based logic components. 

\emph{Acknowledgements}. The authors wish to thank N.M.R. Peres, J.M.B.L.
dos Santos, V.M. Pereira, M.A. Cazalilla, and T.G. Rappoport for helpful
discussions. We also wish to thank V.M. Pereira for pointing us Ref.~\onlinecite{McCann12}.
A.P., A.F. and A.H.C.N. acknowledge support from National Research
Foundation - Competitive Research Programme through grant ``Novel
2D materials with tailored properties: beyond graphene'' (Grant No.
R-144-000-295-281). A.P. and B.O. acknowledge support from Singapore
Millennium Foundation-NUS Research Horizons award (R-144-001-271-592;
R-144-001-271-646). \pagebreak{}

\begin{center}
\textbf{\large Appendix}
\par\end{center}{\large \par}

In this appendix, we rederive the impurity Hamiltonians of Sec.~I
for adatoms in the hollow and top positions, accounting for internal
degrees of freedom. We describe the graphene--adatom system with a
tight-binding Hamiltonian $H=H_{\textrm{gr}}+H_{\textrm{ad}}+V$,
where $H_{\textrm{gr}}$ is pristine graphene's Hamiltonian, $H_{\textrm{ad}}$
is the adatoms' Hamiltonian, and $V$ is the graphene--adatom hybridization
term. We write $H_{\textrm{gr}}$ as the following first-nearest neighbors
tight-binding Hamiltonian: 
\begin{equation}
H_{\textrm{gr}}=-t\sum_{i\in\mathcal{A}}\sum_{\langle i,j\rangle}a_{i}^{\dagger}b_{j}+\textrm{H.c.}\,,
\end{equation}
where $\mathcal{A}$ denotes graphene's $A$-sublattice carbon atoms,
and $a_{i}^{\dagger}$ ($b_{j}$) creates an $A$-sublattice electron
at atom $i\in\mathcal{A}$ (annihilates an electron from the $B$-sublattice
site $j$). Here, $\langle i,j\rangle$ refers to nearest neighbors
\emph{$j$} of site $i$, and \emph{$t$} is the hopping energy between
nearest neighbors. Next, we derive single-electron tight-binding Hamiltonians
for $H_{\textrm{ad}}$ and $V$ by using symmetry arguments,\cite{weeks11}
and then we trace out the adatom degrees of freedom via the Löwdin
transformation.\cite{Lowdin} Taking the continuum-limit then yields
the results of Table \ref{tab:EffectiveMass}. In addition to confirming
the results obtained in Sec.~I, this approach has the advantage of
relating couplings ($V_{0}$, $V_{\textrm{so}}$, $\Delta$, $\Delta_{\textrm{so}}$,
$\Lambda_{R}$, ...) appearing in Table~\ref{tab:EffectiveMass}
to microscopic parameters (hopping integrals, atomic spin--orbit couplings
and energy levels) and the Fermi energy of graphene. 
\begin{figure}[h]
\begin{centering}
\includegraphics[width=0.85\columnwidth]{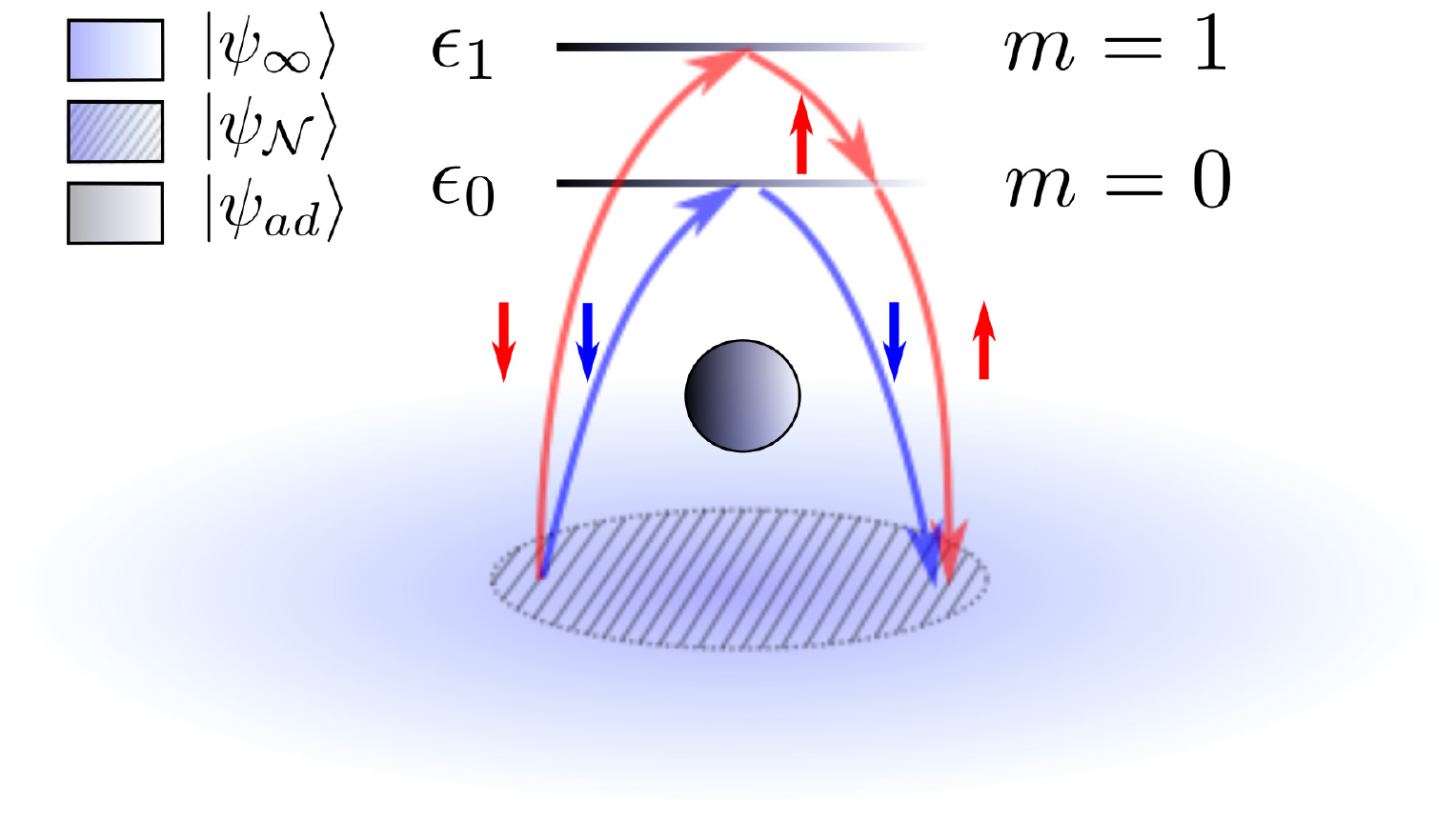}
\par\end{centering}

\caption{\label{fig:Cartoon}Illustration of typical spin-flip (red) and spin-conserving
(blue) processes induced by a $p$-orbital adatom (gray) on graphene
(light blue). Energy levels $\epsilon_{0}$ and $\epsilon_{1}$ of
the adatom's $p$ orbitals $m=0$ and $m=\pm1$ are represented as
gray solid lines. Core orbitals are depicted as a black ball. The
shaded region corresponds to the adatom's immediate vicinity, where
carbon atoms $p_{z}$ orbitals couple strongly to the adatoms valence
$p$ orbital. Red (blue) straight vertical arrows represent the spin
of an electron transiting between graphene and the adatom while flipping
(conserving) its spin. Partial waves $|\psi_{\infty}\rangle$, $|\psi_{\mathcal{N}}\rangle$,
and $|\psi_{\textrm{ad}}\rangle$ introduced in the Appendix are associated
with the blue area, the dashed area, and the adatom's valence orbital,
respectively.}
\end{figure}

We start by writing the solution of the Schrödinger equation $H|\psi\rangle=E|\psi\rangle$
as a sum of waves $|\psi\rangle=|\psi_{\textrm{ad}}\rangle+|\psi_{\mathcal{N}}\rangle+|\psi_{\infty}\rangle$,
where $|\psi_{\textrm{ad}}\rangle$, $|\psi_{\mathcal{N}}\rangle$
and $|\psi_{\infty}\rangle$ are projections of $|\psi\rangle$ on
the adatom valence $l$ orbital, its immediate vicinity---where graphene's
$p_{z}$ orbitals couple strongly to the adatom's valence orbital---and
graphene's distant $p_{z}$ orbitals, respectively (see Fig.~\ref{fig:Cartoon}).
We denote by $d_{m,s}^{\dagger}$ the operator creating an adatom's
$l$ orbital of angular momentum $m$ and spin $s$, and we write
$d_{m,s}^{\dagger}|0\rangle=|m,s\rangle_{\textrm{ad}}$. In the case
of an adatom in the hollow position, $|\psi_{\mathcal{N}}\rangle$
is a linear combination of hexagonal states, $\Omega_{m,s}^{\dagger}|0\rangle=|m,s\rangle_{\mathcal{N}}$.
For top-position adatoms, $|\psi_{\mathcal{N}}\rangle$ is a linear
combination of triangular states, $\Gamma_{m,s}^{\dagger}|0\rangle$
and $c_{0}^{\dagger}|0\rangle$. Here, we explain the method used
in the case of an adatom in the hollow position, the top-position
treatment case being analogous. We write $|\psi_{\textrm{ad}}\rangle$
and $|\psi_{\mathcal{N}}\rangle$ as
\begin{equation}
|\psi_{\mathcal{N}}\rangle=\sum_{s,m}\alpha_{m,s}|m,s\rangle_{\mathcal{N}}\,,
\end{equation}
 
\begin{equation}
|\psi_{\textrm{ad}}\rangle=\sum_{s,m}\beta_{m,s}|m,s\rangle_{\textrm{ad}}\,.
\end{equation}

The projection of the Schrödinger equation $H|\psi\rangle=E|\psi\rangle$
on $|m,s\rangle_{\textrm{ad}}$ gives
\begin{eqnarray}
\sum_{s',m'}\beta_{m',s'}{}_{\textrm{ad}}\langle m,s|H_{\textrm{ad}}|m',s'\rangle_{\textrm{ad}}+\qquad\qquad\qquad\:\:\:\nonumber \\
\sum_{s',m'}\alpha_{m',s'}{}_{\textrm{ad}}\langle m,s|V|m',s'\rangle_{\textrm{\ensuremath{\mathcal{N}}}}=\beta_{m,s}\delta_{m,m^{\prime}}\delta_{s,s^{\prime}}E\,.\label{eq:AdProjection}
\end{eqnarray}
To proceed, we denote by $\hat{\mathcal{Z}}$ the matrix with elements
$\left(_{\textrm{ad}}\langle m,s|H_{\textrm{ad}}|m',s'\rangle_{\textrm{ad}}\right)_{(m,s),(m',s')}$
and by $\hat{\mathcal{T}}$ the matrix with elements $\left(_{\textrm{ad}}\langle m,s|V|m',s'\rangle_{\mathcal{N}}\right)_{(m,s),(m',s')}$.
Within this notation, Eq.~(\ref{eq:AdProjection}) is recast into
the elegant form
\begin{equation}
B=(E\mathbb{I}-\hat{\mathcal{Z}})^{-1}\hat{\mathcal{T}}A\,,\label{eq:BvsA}
\end{equation}
where $A$ and $B$ are vectors with components $\left(\alpha_{m,s}\right)_{(m,s)}$
and $\left(\beta_{m,s}\right)_{(m,s)}$, respectively. Setting $H_{\textrm{imp}}=H_{\textrm{ad}}+V$,
we next project the vector $H_{\textrm{imp}}|\psi\rangle$ on $|m,s\rangle_{\mathcal{N}}$
states. This gives
\begin{equation}
_{\mathcal{N}}\langle m,s|H_{\textrm{imp}}|\psi\rangle=\sum_{m',s'}\alpha_{m',s'}{}_{\mathcal{N}}\langle m,s|\hat{\mathcal{S}}|m',s'\rangle_{\textrm{ad}}\,,\label{eq:NProjection}
\end{equation}
where
\begin{equation}
\hat{\mathcal{S}}=\hat{\mathcal{T}}^{\dagger}(E\mathbb{I}-\hat{\mathcal{Z}})^{-1}\hat{\mathcal{T}}\,.\label{eq:Core}
\end{equation}
Equation~(\ref{eq:NProjection}) can be interpreted as the projection
of the vector $\tilde{H}_{\textrm{imp}}(|\psi\rangle_{\mathcal{N}}+|\psi\rangle_{\infty})$
on the state $|m,s\rangle_{\mathcal{N}}$, where $\tilde{H}_{\textrm{imp}}$
is the \emph{graphene-only} Hamiltonian:

\begin{equation}
\tilde{H}_{\textrm{imp}}=\sum_{m,s}\sum_{m',s'}\hat{\mathcal{S}}_{(m,s),(m',s')}\Omega_{m,s}^{\dagger}\Omega_{m',s'}\,.\label{eq:Graphene-only}
\end{equation}
Tracing $H_{\textrm{ad}}$ out hence consists in replacing $H_{\textrm{imp}}$
by $\tilde{H}_{\textrm{imp}}$ in the full Hamiltonian $H$. 

We now derive a single-electron tight-binding Hamiltonian $H_{\textrm{ad}}$
describing an $l$-orbital adatom either in the hollow or top position,
thereby generalizing a result of Ref.~\onlinecite{weeks11} for $p$
orbitals. We start with an ansatz Hamiltonian $H_{\textrm{ad}}$ that
manifestly conserves total angular momentum, 
\begin{eqnarray}
H_{\textrm{ad}} & = & \sum_{m=-l}^{l}\epsilon_{m}d_{m}^{\dagger}d_{m}+\sum_{m=-l}^{l}\lambda_{\textrm{so}}^{m}d_{m}^{\dagger}s_{z}d_{m}\nonumber \\
 & + & \sum_{m=-l}^{l-1}\Lambda_{\textrm{so}}^{m}(d_{m}^{\dagger}s_{+}d_{m+1}+d_{m+1}^{\dagger}s_{-}d_{m})\,.\nonumber \\
\label{eq:AnsatzHad}
\end{eqnarray}
This Hamiltonian is invariant under rotation by $\pi/3$, so that
choosing energies $\epsilon_{m}$, $\lambda_{\textrm{so}}^{m}$, and
$\Lambda_{\textrm{so}}^{m}$, such that $H_{\textrm{ad}}$ is time-reversal
invariant and symmetric under $\mathcal{R}_{x}:x\mapsto-x$ reflection,
makes it suitable for describing \emph{both} hollow- and top-position
$l$-orbital adatoms, $l=p,d,f$. Since in spherical coordinates,
$\langle\theta,\phi|d_{m}^{\dagger}|0\rangle=Y_{l}^{m}(\theta,\phi)$,
where $Y_{l}^{m}(\theta,\phi)$ are conventional spherical harmonics,
$d_{m}$ transforms into $s_{x}d_{-m}$ under $\mathcal{R}_{x}$,
which sends $\phi$ to $\pi-\phi$. Enforcing $\mathcal{\mathcal{R}}_{x}$-symmetry
thus requires $\epsilon_{m}=\epsilon_{-m}$, $\lambda_{\textrm{so}}^{-m}=-\lambda_{\textrm{so}}^{m}$,
and $\Lambda_{\textrm{so}}^{-m-1}=\Lambda_{\textrm{so}}^{m}$. Moreover,
time-reversal symmetry requires $\epsilon_{m}$, $\lambda_{\textrm{so}}^{m}$,
and $\Lambda_{\textrm{so}}^{m}$ to be reals. We end up with:
\begin{eqnarray}
H_{\textrm{ad}} & = & \sum_{m=-l}^{l}\epsilon_{|m|}d_{m}^{\dagger}d_{m}+\sum_{m=1}^{l}\lambda_{\textrm{so}}^{m}(d_{m}^{\dagger}s_{z}d_{m}-d_{-m}^{\dagger}s_{z}d_{-m})\nonumber \\
 & + & \sum_{m=0}^{l-1}\Lambda_{\textrm{so}}^{m}(d_{m}^{\dagger}s_{+}d_{m+1}+d_{-m-1}^{\dagger}s_{+}d_{-m}+\textrm{H.c.})\,,\nonumber \\
\label{eq:TrueHad}
\end{eqnarray}
which describes the adatom Hamiltonian for both the hollow and top
positions. However, hybridization terms $V$ differ in the hollow-
and top-position cases. We first treat the hollow-position situation,
in which total angular momentum conservation constrains $V$ to take
the form
\begin{align}
V_{\textrm{hollow}} & =\sum_{m=-2}^{2}t_{m}d_{m}^{\dagger}\Omega_{m}+\sum_{m=-2}^{2}\tau_{m}d_{m}^{\dagger}s_{z}\Omega_{m}\nonumber \\
 & +\sum_{m=-2}^{2}(W_{\textrm{so}}^{m}d_{m-1}^{\dagger}s_{+}\Omega_{m}+V_{\textrm{so}}^{m}d_{m+1}^{\dagger}s_{-}\Omega_{m})\nonumber \\
 & +\;\textrm{H.c.}\,.\label{eq:AnsatzHgr-adHollow}
\end{align}
Since $\Omega_{m}\mapsto s_{x}\Omega_{-m}$ under $\mathcal{R}_{x}$,
we must have $t_{m}=t_{-m}$, $\tau_{m}=-\tau_{-m}$, and $W_{\textrm{so}}^{m}=V_{\textrm{so}}^{-m}$.
Enforcing time-reversal symmetry requires $t_{m}$, $\tau_{m}$ and
$W_{\textrm{so}}^{m}$ to read $t_{m}=i^{|m|}u_{|m|}$, $\tau_{m}=i^{m}\nu_{|m|}$,
and $W_{\textrm{so}}^{m}=i^{m}w_{m}$, where $u_{|m|}$, $\nu_{|m|}$,
and $w_{m}$ are real. Finally,
\begin{eqnarray}
V_{\textrm{hollow}} & = & \sum_{m=-2}^{2}i^{|m|}u_{|m|}d_{m}^{\dagger}\Omega_{m}\nonumber \\
 & + & \sum_{m=1}^{2}i^{m}\nu_{|m|}(d_{m}^{\dagger}s_{z}\Omega_{m}-d_{-m}^{\dagger}s_{z}\Omega_{-m})\nonumber \\
 & + & \sum_{m=-2}^{2}i^{m}w_{m}(d_{m-1}^{\dagger}s_{+}\Omega_{m}+d_{-m+1}^{\dagger}s_{-}\Omega_{-m})\nonumber \\
 & + & \textrm{H.c.}\,.\label{eq:TrueHgr-adHollow-1}
\end{eqnarray}
A similar treatment allows us to derive $V$ for top-position adatoms.
Enforcing symmetry under $\mathcal{R}_{x}$, time-reversal symmetry
and total angular momentum conservation, we obtain:
\begin{eqnarray}
V_{\textrm{top}} & = & \sum_{m=-1}^{1}i^{|m|}\theta_{|m|}d_{m}^{\dagger}\Gamma_{m}+i\tau(d_{1}^{\dagger}s_{z}\Gamma_{1}-d_{-1}^{\dagger}s_{z}\Gamma_{-1})\nonumber \\
 & + & \sum_{m=0,1}i^{m}l_{\textrm{so},m}(d_{m-1}^{\dagger}s_{+}\Gamma_{m}+d_{-m+1}^{\dagger}s_{-}\Gamma_{-m})\nonumber \\
 & + & L_{\textrm{so}}(d_{-1}^{\dagger}s_{+}c_{0}+d_{1}^{\dagger}s_{-}c_{0})+\gamma d_{0}^{\dagger}c_{0}+\textrm{H.c.}\,,\label{eq:True_Hgr-ad_top}
\end{eqnarray}
where $\theta_{m}$, $\tau$, $l_{\textrm{so},m}$, $L_{\textrm{so}}$,
and $\gamma$ are real. We can now derive graphene-only Hamiltonians
for adatoms in the hollow or top position using Eqs.~(\ref{eq:Core})
and (\ref{eq:Graphene-only}). We write the $\hat{\mathcal{Z}}$ matrix,
similar for both hollow- and top-position adatoms, in a basis $\mathcal{B}_{l}$
of $2(2l+1)$ states $|m,s\rangle_{\textrm{ad}}$ arranged in ascending
total angular momentum $J=m+s$ order:
\begin{align}
\mathcal{B}_{l} & =\{|-l,\downarrow\rangle_{\textrm{ad}},|-l,\uparrow\rangle_{\textrm{ad}},|-l+1,\downarrow\rangle_{\textrm{ad}},\nonumber \\
 & \hspace{1em}\;\;\;|-l+1,\uparrow\rangle_{\textrm{ad}},...,|l,\downarrow\rangle_{\textrm{ad}},|l,\uparrow\rangle_{\textrm{ad}}\}\,.\label{eq:Bl-basis}
\end{align}

In basis $\mathcal{B}_{l}$, the $\hat{\mathcal{Z}}$ matrix is simply
block-diagonal and reads

\begin{equation}
\hat{\mathcal{Z}}=\left(\begin{array}{ccccc}
\hat{\mathcal{Z}}_{-l-\frac{1}{2}} & 0 & \cdots & 0 & 0\\
0 & \hat{\mathcal{Z}}_{-l+\frac{1}{2}} & \cdots & 0 & 0\\
\vdots & \vdots & \ddots & \vdots & \vdots\\
0 & 0 & \cdots & \hat{\mathcal{Z}}_{l-\frac{1}{2}} & 0\\
0 & 0 & \cdots & 0 & \epsilon_{l}+\lambda_{\textrm{so}}^{l}
\end{array}\right)\,,\label{eq:Z}
\end{equation}
where $\hat{\mathcal{Z}}_{-l-\frac{1}{2}}=\hat{\mathcal{Z}}_{l+\frac{1}{2}}=\epsilon_{l}+\lambda_{\textrm{so}}^{l}$.
If $|J|\neq l+\frac{1}{2}$, then $\hat{\mathcal{Z}}_{J}$ are $2\times2$
matrices acting on total-angular momentum $J$ subspace:
\begin{equation}
\hat{\mathcal{Z}}_{J}=\left(\begin{array}{cc}
E_{J}^{+} & \Delta_{J}\\
\Delta_{J} & E_{J}^{-}
\end{array}\right)\,,\label{eq:Zblock}
\end{equation}
where $E_{J}^{\pm}=\epsilon_{s_{J}(J\mp\frac{1}{2})}\pm s_{J}\lambda_{\textrm{so}}^{s_{J}(J\mp\frac{1}{2})}$,
$\Delta_{J}=\Lambda_{\textrm{so}}^{|J|+\frac{1}{2}}$, $s_{J}=J/|J|$,
and $\lambda_{\textrm{so}}^{0}=0$. Using basis $\mathcal{B}_{\Omega}$
of hexagonal states in ascending-$J$ order,
\begin{align}
\mathcal{B}_{\Omega} & =\{|-2,\downarrow\rangle_{\mathcal{N}},|-2,\uparrow\rangle_{\mathcal{N}},|-1,\downarrow\rangle_{\mathcal{N}},\nonumber \\
 & \quad\:\:\,\:|-1,\uparrow\rangle_{\mathcal{N}},...,|2,\downarrow\rangle_{\mathcal{N}},|2,\uparrow\rangle_{\mathcal{N}}\}\,,\label{eq:BOmega-basis}
\end{align}
the ``hybridization'' matrix $\hat{\mathcal{T}}$ for a hollow-position
adatom is also a sparse matrix. Its only non-zero elements are in
$2\times2$ and $1\times2$ blocks $\hat{\mathcal{T}}_{J}$ connecting
subspaces of hexagonal and adatom orbital states of same total angular
momentum $J$. The $2\times2$ $\hat{\mathcal{T}}_{J}$ blocks read

\begin{equation}
\hat{\mathcal{T}}_{J}=i^{J-\frac{1}{2}}\left(\begin{array}{cc}
a_{J}u_{|J-\frac{1}{2}|}+\nu_{|J-\frac{1}{2}|} & iw_{J+\frac{1}{2}}\\
\\
i^{-2J+1}w_{-J+\frac{1}{2}} & b_{J}u_{|J+\frac{1}{2}|}+i\nu_{|J+\frac{1}{2}|}
\end{array}\right)\,,\label{eq:Tblocks}
\end{equation}
where $a_{J}=i^{|J-\frac{1}{2}|-J+\frac{1}{2}}$ and $b_{J}=i^{|J+\frac{1}{2}|-J+\frac{1}{2}}$,
while $1\times2$ blocks are appropriate sub-matrices of the $2\times2$
blocks shown in Eq.~(\ref{eq:Tblocks}). As a result, the $\hat{\mathcal{S}}$
matrix is block-diagonal and, for instance, the $f$-orbital adatom
graphene-only Hamiltonian reads
\begin{equation}
\tilde{H}_{\textrm{hollow}}=\sum_{m=-l}^{l}(\Omega_{m\uparrow}^{\dagger},\Omega_{m+1\downarrow}^{\dagger})\hat{\mathcal{S}}_{m+\frac{1}{2}}\left(\begin{array}{c}
\Omega_{m\uparrow}\\
\Omega_{m+1\downarrow}
\end{array}\right)\,,\label{eq:SingleElHhollow}
\end{equation}
with 
\begin{equation}
\hat{\mathcal{S}}_{J}=\hat{\mathcal{T}}_{J}^{\dagger}(E\mathbb{I}-\hat{\mathcal{Z}}_{J})^{-1}\hat{\mathcal{T}}_{J}\,.\label{eq:Sblocks}
\end{equation}
Similar results are straightforwardly obtained for $p$- and $d$-orbital
adatoms. The connection with the Hamiltonian of Eq.~(\ref{eq:Hhollow_simple})
is easily made, as
\begin{equation}
\hat{\mathcal{S}}_{m+\frac{1}{2}}=\left(\begin{array}{cc}
\nu_{m}^{+}+\nu_{m}^{-} & i\Lambda_{m}\\
-i\Lambda_{m} & \nu_{m+1}^{+}-\nu_{m+1}^{-}
\end{array}\right)\,.
\end{equation}
We now write the ``hybridization'' matrix $\hat{\mathcal{T}}_{\textrm{top}}$
for top-position adatoms, using a basis $\mathcal{B}_{\textrm{top}}$
of states arranged in ascending-$J$ order:
\begin{align}
\mathcal{B}_{\textrm{top}} & =\{\Gamma_{-1,\downarrow}^{\dagger}|0\rangle,\Gamma_{-1,\uparrow}^{\dagger}|0\rangle,\Gamma_{0,\downarrow}^{\dagger}|0\rangle,c_{0,\downarrow}^{\dagger}|0\rangle,\nonumber \\
 & \quad\:\, c_{0,\uparrow}^{\dagger}|0\rangle,\Gamma_{0,\uparrow}^{\dagger}|0\rangle,\Gamma_{1,\downarrow}^{\dagger}|0\rangle,\Gamma_{1,\uparrow}^{\dagger}|0\rangle\}\,.\label{eq:Btop}
\end{align}
Unlike $\mathcal{B}_{\Omega}$, $\mathcal{B}_{\textrm{top}}$ comprises
six states of total angular momentum $\pm1/2$ due to the presence
of central states $c_{0,\uparrow/\downarrow}^{\dagger}|0\rangle$
in addition to triangular states $\Gamma_{0,\uparrow/\downarrow}^{\dagger}|0\rangle$
. Correspondingly, the only non-zero elements of $\hat{\mathcal{T}}_{\textrm{top}}$
are in blocks $\hat{\mathcal{T}}_{\textrm{top},J}$ connecting states
of total angular momentum $J$. Irrespective of the adatom's valence
orbital---$p$, $d$, or, $f$---we have
\begin{equation}
\hat{\mathcal{T}}_{\textrm{top},-\frac{1}{2}}=\left(\begin{array}{ccc}
i\theta_{-1}+i\tau & l_{\textrm{so},0} & L_{\textrm{so}}\\
il_{\textrm{so},1} & \theta_{0} & \gamma
\end{array}\right)\,,
\end{equation}
\begin{equation}
\hat{\mathcal{T}}_{\textrm{top},\frac{1}{2}}=\left(\begin{array}{ccc}
\gamma & \theta_{0} & il_{\textrm{so},1}\\
L_{\textrm{so}} & l_{\textrm{so},0} & i\theta_{1}-i\tau
\end{array}\right)\,.
\end{equation}
Other blocks $\hat{\mathcal{T}}_{\textrm{top},J}$ depend on the adatom's
valence orbital, but are appropriate sub-arrays of
\begin{equation}
\hat{\mathcal{T}}_{\textrm{top},-\frac{3}{2}}=\left(\begin{array}{c}
-il_{\textrm{so},-1}\\
i\theta_{1}+i\tau
\end{array}\right)\,,
\end{equation}
\begin{equation}
\hat{\mathcal{T}}_{\textrm{top},\frac{3}{2}}=\left(\begin{array}{c}
i\theta_{1}+i\tau\\
-il_{\textrm{so},-1}
\end{array}\right)\,.
\end{equation}
The $\hat{\mathcal{S}}$ matrix for top-position adatoms is block-diagonal,
\begin{equation}
\hat{\mathcal{S}}_{\textrm{top}}=\left(\begin{array}{cccc}
\hat{\mathcal{S}}_{\textrm{top},-\frac{3}{2}} & 0 & 0 & 0\\
0 & \hat{\mathcal{S}}_{\textrm{top},-\frac{1}{2}} & 0 & 0\\
0 & 0 & \hat{\mathcal{S}}_{\textrm{top},\frac{1}{2}} & 0\\
0 & 0 & 0 & \hat{\mathcal{S}}_{\textrm{top},\frac{3}{2}}
\end{array}\right)\,,
\end{equation}
with $\hat{\mathcal{S}}_{\textrm{top},J}=\hat{\mathcal{T}}_{\textrm{top},J}^{\dagger}(E\mathbb{I}-\hat{\mathcal{Z}}_{J})^{-1}\hat{\mathcal{T}}_{\textrm{top},J}$
as in Eq.~(\ref{eq:Sblocks}). The graphene-only top-position Hamiltonian
thus reads
\begin{eqnarray}
\tilde{H}_{\textrm{top}} & = & \hat{\mathcal{S}}_{\textrm{top},-\frac{3}{2}}\Gamma_{-1,\downarrow}^{\dagger}\Gamma_{-1,\downarrow}+\hat{\mathcal{S}}_{\textrm{top},\frac{3}{2}}\Gamma_{1,\uparrow}^{\dagger}\Gamma_{1,\uparrow}\nonumber \\
 & + & (\Gamma_{-1,\uparrow}^{\dagger},\Gamma_{0,\downarrow}^{\dagger},c_{0,\downarrow}^{\dagger})\hat{\mathcal{S}}_{\textrm{top},-\frac{1}{2}}\left(\begin{array}{c}
\Gamma_{-1,\uparrow}\\
\Gamma_{0,\downarrow}\\
c_{0,\downarrow}
\end{array}\right)\nonumber \\
 & + & (c_{0,\uparrow}^{\dagger},\Gamma_{0,\uparrow}^{\dagger},\Gamma_{1,\downarrow}^{\dagger})\hat{\mathcal{S}}_{\textrm{top},\frac{1}{2}}\left(\begin{array}{c}
c_{0,\uparrow}\\
\Gamma_{0,\uparrow}\\
\Gamma_{1,\downarrow}
\end{array}\right)\,.\label{eq:SingleElHtop}
\end{eqnarray}
This is exactly the Hamiltonian of Eq.~(\ref{eq:Htop}) with the
following correspondence
\begin{equation}
\hat{\mathcal{S}}_{\textrm{top},-\frac{3}{2}}=\hat{\mathcal{S}}_{\textrm{top},\frac{3}{2}}=\Lambda_{+}+\Lambda_{-}\,,\label{eq:Stop_1p5}
\end{equation}
\begin{equation}
\hat{\mathcal{S}}_{\textrm{top},-\frac{1}{2}}=\left(\begin{array}{ccc}
\Lambda_{+}-\Lambda_{-} & -i\tau & -i\mu\\
i\tau & V_{1} & V_{2}\\
i\mu & V_{2} & V_{0}
\end{array}\right)\,,\label{eq:Stop_m0p5}
\end{equation}
and
\begin{equation}
\hat{\mathcal{S}}_{\textrm{top},\frac{1}{2}}=\left(\begin{array}{ccc}
V_{0} & V_{2} & i\mu\\
V_{2} & V_{1} & i\tau\\
-i\mu & -i\tau & \Lambda_{+}-\Lambda_{-}
\end{array}\right)\,.\label{eq:Stop_0p5}
\end{equation}
\bigskip{}

\end{document}